\documentclass[pdflatex,sn-aps]{sn-jnl}

\usepackage{graphicx}%
\usepackage{multirow}%
\usepackage{amsmath,amssymb,amsfonts}%
\usepackage{amsthm}%
\usepackage{mathrsfs}%
\usepackage[title]{appendix}%
\usepackage{xcolor}%
\usepackage{textcomp}%
\usepackage{manyfoot}%
\usepackage{booktabs}%
\usepackage{algorithm}%
\usepackage{algorithmicx}%
\usepackage{algpseudocode}%
\usepackage{listings}%

\theoremstyle{thmstyleone}%
\theoremstyle{thmstyletwo}%

\theoremstyle{thmstylethree}%

\usepackage{xcolor}
\usepackage{graphicx}
\usepackage{amsmath}
\usepackage{braket}
\usepackage{tikz}
\usepackage{accents}
\usepackage{ulem}
\definecolor{nblue}{rgb}{0.06,0.3,0.73}
\definecolor{nblack}{rgb}{0,0,0}
\definecolor{nred}{rgb}{0.9,0.1,0.1}
\definecolor{nmagenta}{rgb}{0.7,0.0,0.3}
\definecolor{npurple}{rgb}{0.52,0,0.52}
\definecolor{neditcolor}{rgb}{0.3,0.3,0.9}
\definecolor{colq1}{HTML}{FC0D1B}

\newcommand{\red}{\color{nblack}}

\newcommand{\blk}{\color{nblack}}
\usepackage{bm}
\def\be{\begin{equation}}
\def\ee{\end{equation}}
\def\la{\langle}
\def\ra{\rangle}
\newcommand\norm[1]{||#1||}
\newcommand{\n}{\nabla}
\newcommand{\bb}[1]{\boldsymbol{ #1}}  
\newcommand{\Tr}{\text{Tr}}
\newcommand{\dg}{^\dagger}
\newcommand{\dd}{{\rm d}}
\newcommand{\dt}{\dd t}
\newcommand{\ddt}{\delta t}

\newcommand{\accentbotharrowc}[1]{%
	\begin{tikzpicture}[#1]%
		\fill (-0.6mm,0) -- (0.2mm,0.3mm) -- (0.2mm,-0.3mm);
		\fill (2mm,0) -- (1.2mm,0.3mm) -- (1.2mm,-0.3mm);
		\draw[line width = 0.2mm] (-0.1mm,0mm) -- (1.2mm,0mm);
		\draw (2mm,0) -- (1.2mm,0.3mm) -- (1.2mm,-0.3mm) -- cycle;
		\draw (-0.6mm,0) -- (0.2mm,0.3mm) -- (0.2mm,-0.3mm) -- cycle;
	\end{tikzpicture}%
}
\newcommand{\tempbothc}{\accentbotharrowc{}}
\newcommand{\bothp}[1]{\accentset{\tempbothc}{#1}}

\newcommand{\accentbotharrowo}[1]{%
	\begin{tikzpicture}[#1]%
		\fill (-0.6mm,0) -- (0.2mm,0.3mm) -- (0.2mm,-0.3mm);
		\draw[line width = 0.2mm] (-0.1mm,0mm) -- (1.2mm,0mm);
		\draw (2mm,0) -- (1.2mm,0.3mm) -- (1.2mm,-0.3mm) -- cycle;
		\draw (-0.6mm,0) -- (0.2mm,0.3mm) -- (0.2mm,-0.3mm) -- cycle;
	\end{tikzpicture}%
}
\newcommand{\tempbotho}{\accentbotharrowo{}}
\newcommand{\both}[1]{\accentset{\tempbotho}{#1}}

\begin{document}

\title[Quantum state-preparation control in noisy environment via most-likely paths]{Quantum state-preparation control in noisy environment via most-likely paths}


\author[1,2]{\fnm{Wirawat} \sur{Kokaew}}\email{wkokaew@perimeterinstitute.ca}

\author*[1,2]{\fnm{Thiparat} \sur{Chotibut}}\email{thiparat.c@chula.ac.th}

\author*[3]{\fnm{Areeya} \sur{Chantasri}}\email{areeya.chn@mahidol.ac.th}

\affil[1]{\orgdiv{Department of Physics, Faculty of Science}, \orgname{Chulalongkorn University}, \orgaddress{\street{Phaya Thai Road}, \city{Bangkok}, \postcode{10330}, \country{Thailand}}}

\affil[2]{\orgdiv{Chula Intelligent and Complex Systems Research Unit, Faculty of Science}, \orgname{Chulalongkorn University}, \orgaddress{\street{Phaya Thai Road}, \city{Bangkok}, \postcode{10330}, \country{Thailand}}}

\affil[3]{\orgdiv{Department of Physics, Faculty of Science}, \orgname{Mahidol University}, \orgaddress{\street{Rama VI Road}, \city{Bangkok}, \postcode{10400}, \country{Thailand}}}


\abstract{
Finding controls for open quantum systems needs to take into account effects from unwanted environmental noise. Since actual realizations or states of the noise are typically unknown, the usual treatment for the quantum system's decoherence dynamics is via the so-called Lindblad master equation, which in essence describes an average evolution (mean path) of the system's state affected by the unknown noise. We here consider an alternative view of a noise-affected open quantum system, where the average dynamics can be unravelled into hypothetical noisy quantum trajectories, and propose a control strategy for the state-preparation problem based on the likelihood of noise occurrence. We formulate a stochastic path integral for noise variables whose extremum yields control functions associated with a most‑likely noise to achieve target states. As a proof of concept, we apply our method to a qubit-state preparation under dephasing noise and analytically solve for controlled Rabi drives for arbitrary target states. Since the method is constructed based on the probability of noise, we also introduce a fidelity success rate as a measure of the state preparation. We benchmark against the mean‑path approaches, e.g., GRAPE and CRAB controls, using both average fidelity and a success‑rate metric. While standard mean‑path controls maximize average fidelity, most-likely controls achieve higher success rates, especially at strong dephasing. 
}




\maketitle

\section{Introduction}
The need for controls of open quantum systems~\cite{Koch2016,BookWiseman,DonPet2022,Carrie2025}~stems from the fact that perfectly closed quantum systems have been unachievable in real experiments and environmental noises always affect dynamics of the systems.
Various techniques for quantum control which initially assumed unitarily evolving quantum systems have therefore been modified to include dissipative or noise disturbance effects. Examples are Gradient-based optimal controls using quantum trajectories~\cite{GoeJac2018,AbdSch2019}, Pontryagin’s maximum principle for open quantum systems~\cite{SugKon2007,CavMar2018,LinSel2020,BosSig2021,LewWha2022}, and, more recently, machine learning-assisted quantum controls~\cite{NiuBoi2019,YouPaz2020,LuiPie2022,HuaBan2022,Sivak2022,Dong2023}.
One of the most common modifications is to replace the system's dynamical equation by the Gorini-Kossakowski-Lindblad-Sudarshan  equation~\cite{GKS1976,Lind1976,ChrPas2017}, typically known as Lindblad master equation. 
Solutions of the Lindblad equation are extensively used in explaining decoherence effects, such as relaxation and pure dephasing, occurring in experiments~\cite{ChiBur2008,Sch2019}.



However, the Lindblad equation describes a reduced system's dynamics from considering system-environment interactions and tracing out the environmental part~\cite{BookDavies,BookCarmichael2,BookBreuer}. Interestingly, if one considers the traced-out environmental part as actually existing in a particular state, simply unknown to us, then it can be shown that a solution of the Lindblad master equation does represent an average path or a ``mean estimator" of the unknown system's evolutions associated with possible realizations of the environmental states. 
Thus, from the estimation theory's point of view, the Lindblad solution is not the only representative of the system's state dynamics, but rather one of the many possible choices including the most-likely or median trajectory estimators~\cite{ChaGue2021}.   
In this work, in the context of controlling open quantum systems undergoing interaction with noisy environment, our aim is to explore this alternative view that the noisy environment can be \textit{unravelled} into ensembles of possible states and their corresponding system's trajectories~\cite{Davies1969,BookCarmichael,BookBachielli,Wiseman1996,BookJacobs,Jacobs2006}, and use the latter as the representatives of the system's dynamics in order to explore a new approach for quantum controls based on trajectory probabilities.

In particular, we consider the state-preparation control problem, which is the subject of recent research interest~\cite{ComWis2011,ZhaWei2019,GunPet2021,PorEss2022,Perret2024}, where the task is to optimally control an open quantum system from its initial state to a (pure) target state under the disturbance of an environmental noise. With the conventional Lindblad mean-path (MP) approach, one can solve the master equation for different control functions and search for the control that gives the highest fidelity between the final state and the target. However, since the Lindblad solution describes an average of the system's evolutions conditioned on environmental states, its prediction for a final system's state will typically be a mixed state. Thus maximising the fidelity between the target state and the final state is equivalent to maximising an average of fidelities between the target state and all possible unravelled final states.
This latter argument does raise an interesting question whether we could consider other objective functions to be maximised in the state-preparation control.

Instead of maximising the average fidelity, we here consider maximising the chance to achieve the near-unit fidelity given the noisy environment. We propose the use of the most-likely path (MLP) introduced in \cite{Chantasri2013,chantasri2015stochastic}. The MLP approach, originally proposed for continuously monitored quantum trajectories, utilizes a joint probability density function of noisy trajectories expressed in the form of a stochastic path integral, where its action can be extremised to solve for the least-action path between any two boundary states. We modify the original MLP approach by replacing the monitored quantum trajectories with the unravelled noisy trajectories and adding control functions in the action to be extremised. This extremised-action solution therefore gives the control function and the system's path that optimise the probability of noise to reach the target state~\cite{chantasri2015stochastic,LewCha17}. This MLP technique, even though derived independently from the stochastic path integral, could also be viewed as a Pontryagin-like optimal control, where the optimality condition is the probability density function of the unravelled noise~\cite{BosSig2021,LewWha2022}. Therefore, in order to benchmark our approach against the conventional MP one, we then introduce a new measure, \textit{success rate}, to represent the probabilities of the controlled system's final states being in proximity to the target state.

As a proof of concept, we apply the MP and MLP techniques to a toy model of preparing a qubit state under a typical pure dephasing noise, where only a single Rabi drive with controllable strength is allowed. For the MP approach, the Lindblad solution has to be solved with numerical optimisation tools such as the Gradient Ascent Pulse Engineering (GRAPE) and the Chopped RAndom Basis (CRAB)\cite{Khaneja2005,Caneva2011,Goerz2019,Johansson}, while the MLP optimal Rabi drive can be solved analytically for arbitrary target states, which is considered as an advantage over the MP method.
We investigate the MP and MLP control performances by analysing average fidelities, the distribution of controlled final states, and the success rates, to various choices of target states. We find that, despite offering less average fidelity than the MP controls, the optimal controls from the MLP approach can surprisingly result in higher success rates in reaching particular targets. We present numerical results and analyse case-by-case when the MLP controls can outperform the MP ones or vice versa. We also note that this single-qubit control example, even though seemed trivial, could be used as the first step towards building control algorithms for more sophisticated models.

The paper is organized as follows. In Section~\ref{sec2}, we briefly review the Lindblad (mean-path) approach for controlling an open quantum system and then propose a concept of quantum control for unravelled quantum trajectories. In Section~\ref{sec3}, we introduce the stochastic path integral formalism and the most-likely path approach adapted for the state-preparation quantum control, where the success rate is defined as a quality metric. We then apply the control approaches to the example of a two-level system coupled to a pure-dephasing environment, analytically finding the optimal Rabi drives in Section~\ref{sec4}. Numerical simulations and analyses are provided in Section~\ref{sec5}, where the MP and MLP approaches are benchmarked against each other for different target states, decoherence rates, as well as for single/multi-pulse Rabi controls. The conclusion can be found in Section~\ref{sec6} and the full derivations and detailed discussions of the numerical simulations are presented in the Appendices.

\section{Lindblad evolution and its unravelling for quantum control}\label{sec2}

\subsection{Lindblad ``mean-path" approach for quantum control}\label{sec-mp-app}
In the past decade, problems in quantum control have been generalized to include effects from unwanted environmental noise. Quantum systems of interest can no longer be treated as closed systems with pure unitary dynamics, but rather be treated as \textit{open} quantum systems subject to effects from measurements and decoherence.  
The conventional treatment of an open quantum system is to construct a combined system, which consists of the system of interest (S) and its environment or bath (B), described by a total Hamiltonian $\hat H_{\rm SB}(t)$. The combined state of the system of interest and the bath, denoted by $\varrho(t)$, is assumed to evolve unitarily as,
\be\label{eq-combine}
\varrho(t) = \hat U_{\rm SB}(t,0) \varrho(0) \hat U_{\rm SB}(t,0)\dg,
\ee
where the unitary operator is given by, $\hat U_{\rm SB}(t,0) \equiv \exp[-i \! \int_0^t \dd s \, \hat H_{\rm SB}(s)]$. To describe the system-bath interaction, the Hamiltonian $\hat H_{\rm SB}(s)$, can include an interaction term of the form $\sum_k \hat L_k \otimes \hat B_k$, for the system's operators $\hat L_k$ and the bath's operators $\hat B_k$. In the case when only the system's part is of interest, one can trace out the environmental degrees of freedom to obtain the reduced dynamics of the system's state alone. That is, the system's state is given by, $\rho(t) \equiv \rho_{\rm S}(t) = {\rm Tr}_{\rm B}[\varrho(t)]$. Under the strong Markov assumption \cite{LiLi2018}, the reduced dynamics can be generally described by the Lindblad master equation,
\begin{equation}\label{Lindblad}
\dd \rho(t) =- i \, \dt [\hat{H},\rho(t)] + \dt \sum^K_{k=1}\mathcal{D}[\hat{L}_k]\rho(t),
\end{equation}
where the first term describes any unitary part of the evolution generated by a Hermitian operator $\hat{H}$ and the second term describes non-unitary decoherence effects from the system-bath interaction. The superoperator in the second term is defined as $\mathcal{D}[\hat{L}_k]\bullet\equiv\hat{L}_k\bullet \hat{L}_k^\dagger -(\hat{L}_k^\dagger \hat{L}_k\bullet+\bullet \hat{L}_k^\dagger \hat{L}_k)/2$, where $\{\hat{ L}_k\}$ is the set of the Lindblad operators describing all the system-bath coupling channels.

The aim of quantum control for open quantum systems is to engineer the unitary and non-unitary dynamics in Eq.~\eqref{Lindblad} such that the system's state $\rho(t)$ behaves as desired. The most common strategy is to add external linear unitary controls $\hat H_{\rm ctrl} = \sum_{j=1}^J u_j(t)\hat{L}_j$ to the system, where $u_j(t)$'s are control functions associated with the system's operators $\hat L_j$'s. The system's state evolution then becomes
\begin{equation}\label{Lindblad-ctrl}
\dd \rho(t) =-i \, \dt \Big[\hat{H}+\sum_{j=1}^J u_j(t)\hat{L}_j,\rho(t)\Big]+\dt \sum^K_{k=1}\mathcal{D}[\hat{L}_k]\rho(t) .
\end{equation}
The control functions, $u_j(t)$'s, can then be optimised with respect to an objective function, which is to be specifically defined. In most cases, the control problems of this kind cannot be solved analytically and solutions can only be found via numerical optimisations. Several numerical methods for quantum control have been proposed, such as the Gradient Ascent Pulse Engineering (GPAPE) and the Chopped RAndom Basis (CRAB). GRAPE is often used in finding optimal sequence pulses and has been successfully applied in nuclear magnetic resonance systems~\cite{Khaneja2005}. On the other hand, CRAB uses Fourier coefficients and is typically applied to problems with large parameter space~\cite{Caneva2011}. 

\subsection{Maximising fidelity for state preparation}

Let us now consider a specific control problem: state preparation for open quantum systems. The main task is to prepare a quantum system of interest in a desired target state $\rho_{\rm F}$, at a particular time $T$, where the system's dynamics at other intermediate times are irrelevant. In this case, a control objective shall indicate the difference between the actual state \red at the final time\blk, $\rho(T)$, and the \red ``final" \blk target state $\rho_{\rm F}$. The most ubiquitous choice of an objective function is the \textit{quantum state fidelity}~\cite{Jozsa1994} defined as
\begin{equation}\label{eq-fidel}
    \mathcal{F}[\rho(T),\rho_{\rm F}] \equiv \left\{ {\rm Tr}\left[\sqrt{{\rho_{\rm F}}^{1/2}\rho(T){\rho_{\rm F}}^{1/2}}\right] \right\}^2 = {\rm Tr}\left[ \rho(T)\, \rho_{\rm F}\right],
\end{equation}
for the two state matrices. We note that the last equality holds only when at least one of the two states is pure, which is typically satisfied because the target state is usually a pure state, i.e., ${\rm Tr}[\rho_{\rm F}^2]=1$. Given this objective function, one can then use the numerical methods mentioned earlier to solve for optimal control functions, $u_j(t)$ in Eq.~\eqref{Lindblad-ctrl}, that maximise the fidelity in  Eq.~\eqref{eq-fidel}. 



\subsection{Unitary unravelling of a Lindblad evolution}\label{sec-unravel}

We emphasize that the dynamics of open quantum systems in Eq.~\eqref{Lindblad} is obtained by tracing out all the environmental degrees of freedom. Therefore, \red if somehow the state or information of the environment can be revealed, \blk the Lindblad evolution can be considered unravelled into many possible system's evolutions, where each evolution is conditioned on a particular realization of the environmental state. The unravelling process can be applied \red hypothetically by assuming {\it a priori} statistical properties of the environmental state, e.g., see Ref.~\cite{GoeJac2018} for the use of unravelled trajectories for quantum control or Refs.~\cite{Ivonne2019,chantasri2019,ChaGue2021} for the use of unravelled trajectories for quantum state smoothing. The statistical properties can also be obtained more accurately by using the spectral density information~\cite{BookGardiner,YugSas2011,YouWha2012,PazVio2014} of the environment (as random processes are related to their spectra via Fourier transformation). The unravelling can be even more accurate, if there are measurements with known outcomes that partially reveal some information about the environment, such as in Refs.~\cite{WuTur2021,TurWu2022}, or fully reveal the meter's states (e.g., microwave probing fields) as in Refs.~\cite{murch2013observing,ibarcq2015,jordan2015fluores,Shay2016noncom}. \blk

We are interested in a problem of quantum control where the Lindblad evolution can be unravelled based on statistical properties of the environment \red and any known information from measuring the environment. Let us consider a single decoherence channel, $K=1$ in Eq.~\eqref{Lindblad-ctrl}, for simplicity. \red We first remove any fast system-only and environment(bath)-only Hamiltonians by moving into a rotating frame that evolves with them. The system-bath dynamics can then be described by a time-dependent Hamiltonian,
\be\label{eq-hamtot}
\hat H_{\rm SB}(t) = \hat H_{\rm S} \otimes \hat I_{\rm B} + g\, \hat S\red(t)\blk \otimes \hat B(t),
\ee
where the first term is a leftover system-only Hamiltonian, denoted by $\hat H_{\rm S}$, in tensor product with the identity of the bath's Hilbert space, $ \hat I_{\rm B}$. \red
 The second term of Eq.~\eqref{eq-hamtot} is the coupling Hamiltonian where a system's observable, $\hat S(t)$, is coupled to a time-dependent bath's observable, $\hat B(t)$, with $g$ being \blk the system-bath coupling strength. We note that, in the following bath-unravelling derivation, we model the environmental noise to fluctuate on a timescale of $\delta t$ and its values at different $\delta t$-interval are determined by collapses of bath's (mixed) states, $\rho_{\rm B}(t)$, to eigenstates of the bath's observable $\hat B(t)$. To some expert readers, this quantum bath's model simply describes a classical Markovian noise; however, we would like to keep this general framework as it can facilitate the use of Lindblad unravelling for other non-classical environmental noises~\cite{PazNor2017,ChaTon2021}\blk. 

Using the Born-Markov approximation, when the bath is \red assumed large enough\blk, the bath's state changes slowly compared to the rapid system's dynamics. \red That is, if the bath's state changes with a timescale of $\delta t$, then the system-bath combined state in Eq.~\eqref{eq-combine} can be approximately factorized as $\varrho(t') \approx \rho(t') \otimes \rho_{\rm B}(t)$ for $t' \in [t, t+\delta t)$\blk. Also, the bath's observable is assumed to be uncorrelated, i.e., ${\rm Tr}[\hat B(t_1) \hat B(t_2) \rho_{\rm B}] = \kappa \delta(t_1-t_2)$, where $t_1$ and $t_2$ are times in different $\delta t$ intervals and \blk $\kappa$ is the noise spectral density. We show in Appendix \ref{app-open} that, with the Hamiltonian in Eq.~\eqref{eq-hamtot}, tracing out the environmental degrees of freedom from the evolution of the combined system in Eq.~\eqref{eq-combine} leads to the following system-only dynamics \red during the bath's timescale $\delta t$,
\be\label{eq-decoh}
 \rho(t+\delta t) = e^{{\cal L}\delta t}\rho(t),
\ee
where the Lindblad superoperator $\cal L$ is given by
\be
{\cal L}\rho(t) \equiv -i \,  [\hat{H}_{\rm S} ,\rho(t) ]+\gamma  \, \mathcal{D}[\hat S]\rho(t) ,
\ee\blk
with  $\gamma = 2 g^2 \kappa$. The second term is the decoherence effect resulting from the system-environment interaction.

To unravel the Lindblad evolution of Eq.~\eqref{eq-decoh}, we first assume that the bath's state, $\rho_{\rm B}(t)$, \red at the beginning of each time interval, $t' \in [t,t+\delta t)$\blk, can be written as a mixed state on the instantaneous eigenstates of the observable $\hat B(t)$, i.e.,
\be\label{eq-bathprob}
\rho_{\rm B}(t) =\int \dd \xi(t)\,  \wp[\xi(t)] \,  |\xi(t)\ra\la \xi(t)|,
\ee 
where $\hat B(t) |\xi(t)\ra = \xi(t) |\xi(t)\ra$, the integration is over all possible values of $\xi(t)$ at time $t$, and the probability density function $\wp[\xi(t)]$  is normalized as $\int \dd \xi(t) \wp[\xi(t)] = 1$. \red It is important to note here that the function $\wp[\xi(t)]$ can be determined based on any known information about the bath's state $\rho_{\rm B}(t)$. For example, for a thermal bath, the distribution of its quadrature's eigenvalue is Gaussian~\cite{BookGardiner}. Moreover, if there is additional information, denoted by $R$, obtained from bath's measurements, one can construct a conditional probability density function, $\wp[\xi(t)|R]$, to be used in place of $\wp[\xi(t)]$ in the above equation.

Now, suppose the environment is hypothetically \blk found in one of the possible states, $|\xi(t)\ra$, we can calculate the system's state conditioned on that finding by projecting the bath's state onto the eigenstates before tracing out \red to remove \blk the bath's degrees of freedom. That is, given an arbitrary system-bath state $\varrho(t) \red = \rho(t) \otimes \rho_{\rm B}(t)$ at time $t$, we can compute a conditional system's state, at an infinitesimal time $\ddt$ later, from
\begin{align}\label{eq-unrav}
\rho_{\xi}&(t+\ddt) \propto  {\rm Tr}_{\rm B}\left\{ |\xi(t)\ra\la \xi(t)| \hat U_{\rm SB}(t+\ddt,t) \, \varrho(t) \, \hat U_{\rm SB}\dg(t+\ddt,t) \right\}.
\end{align}
By substituting the system-bath Hamiltonian of Eq.~\eqref{eq-hamtot} and using properties of eigenstates \red in Eq.~\eqref{eq-bathprob}, we find that
\be\label{eq-unrav2}
\rho_{\xi}(t+\ddt) \propto  \exp[-i \hat H_{\xi}(t) \ddt] \rho(t) \exp[+i \hat H_{\xi}(t) \ddt],
\ee
where we have defined the \textit{unravelled Hamiltonian} as
\be\label{eq-stoHamil}
\hat H_{\xi} = \hat H_{\rm S}+ g\, \xi(t) \hat S.
\ee
We can also calculate the normalization factor in both Eq.~\eqref{eq-unrav} and Eq.~\eqref{eq-unrav2} from
\begin{align}\label{eq-probweight}
& {\rm Tr}_{\rm B, S}\bigg\{ |\xi(t)\ra\la \xi(t)| \hat U_{\rm SB} (t+\ddt,t) \, \varrho(t) \, \hat U_{\rm SB}\dg(t+\ddt,t) \bigg\} \nonumber \\
 &= \la \xi(t) | \rho_{\rm B}(t) | \xi(t)\ra = \wp[\xi(t)],
\end{align}
which is the probability density of finding the bath's state in the eigenstate $|\xi(t)\ra$ defined in Eq.~\eqref{eq-bathprob}. From the unravelled process of Eq.~\eqref{eq-unrav}, we can see that averaging the conditional state over all possible bath's eigenstates with the probability weight of Eq.~\eqref{eq-probweight} gives back the simple trace over the bath state space, i.e.,
\begin{align}\label{eq-averunrav}
\rho(t+\delta t) \,\,=& \int \! \dd \xi(t) \, \wp[\xi(t)] \, \rho_{ \xi}(t+\ddt) \nonumber \\
=& \int \! \dd \xi(t) \, \la \xi(t)| \hat U_{\rm SB} (t+\ddt,t) \, \varrho(t) \, \hat U_{\rm SB}\dg(t+\ddt,t)|\xi(t)\ra \nonumber \\
=&  \,\, e^{{\cal L}\ddt}\rho(t)
\end{align}
which is the system's state evolution \red described in Eq.~\eqref{eq-decoh}.

\red In the case where the total evolution time is much longer than the bath's fluctuation timescale, we can further assume for simplicity that \blk $\delta t \rightarrow \dt$ and expand Eq.~\eqref{eq-unrav2} to the first order in $\dt$ to obtain the \textit{unravelled} system's dynamics
\be\label{eq-unravelk}
\dd \rho_{\xi}(t)=-i \, \dt [ \hat{H}_{\xi}(t),\rho_{\xi}(t) ].
\ee
We can see that the system evolution in Eq.~\eqref{eq-decoh} is now unravelled into many possible \textit{unitary evolutions} described by the Hamiltonian, $\hat H_{\xi}(t) = \hat H_{\rm S}+ g\, \xi(t) \hat S$, conditioned on the particular environmental (bath) dynamics, $\xi(t)$. We note that the calculation above, even though limited to the unitary unravelling (with a Hermitian $\hat S$), can be applied to a variety of noisy dynamics of $\xi(t)$, including the dynamics induced by classical noises such as thermal noises, fluctuations in classical fields, or fluctuations in any environmental parameters.



\subsection{Quantum control with unravelled evolutions and objectives on state fidelity}\label{sec-avefid}

Quantum control for open quantum systems can be designed differently based on the information of the environment. If the environmental state is known via measurement, a measurement-feedback regime can be implemented, where the measurement results are taken into account when searching for optimal controls. If there is no measurement, the Lindblad evolution can be unravelled based on statistical properties of the bath, such as types and probability distributions of the bath's states that affect the system.

Given the example of unitarily unravelling the Lindblad evolution in the previous subsection, we can consider adding quantum controls to the unravelled state trajectory, similar to Eq.~\eqref{Lindblad-ctrl}, which leads to the dynamical equation,
\be\label{eq-unravelk1}
\dd \rho_{\xi}(t)=-i \, \dt \Big[ \hat{H}_{\xi}(t)+\sum_{j=1}^J u_j(t)\hat{L}_j ,\rho_{\xi}(t) \Big],
\ee
where the control functions $u_j(t)$'s are chosen to optimise an objective function defined for a particular problem. Considering the quantum state-preparation problem, the typical objective function is the fidelity between the target state and the system's state at the final time $T$. In the case when the noise realization $\xi(t)$ is unknown, then the standard approach is to use the ``average" dynamics or the ``mean path," $\rho(t)$, in Eq.~\eqref{eq-decoh}, as the representative of the quantum state evolution and then search for optimal control functions that maximise the fidelity between the target state and the average state $\rho(T)$.

However, here we analyze that the mean-path approach, using the fidelity to the average final state as the objective function, is not the only approach for the dynamics that can be unravelled by the environmental noise state. Let us discretise the time into $t \in \{ 0, \ddt, 2\ddt, ..., T\}$ \red using the bath's timescale, $\delta t$\blk, and write the single time-realization of the bath state $|\xi(t)\ra$ for $t \in [0,T]$ as $|\xi_0\ra, |\xi_{\ddt}\ra,..., |\xi_{T-\ddt}\ra$. From the average dynamics in Eq.~\eqref{eq-averunrav} and the unravelled dynamics in Eq.~\eqref{eq-unrav2}, we can show that the average state \red at the final time \blk is
\begin{align}\label{eq-finalave}
\rho(T)  =\, e^{{\cal L}T}\rho(0) = \int \! \prod_{t=0}^{T-\ddt} \dd \xi_t \,  \wp[\xi_0,..., \xi_{T-\ddt}] \, {\cal U}_{\xi(T)} \rho(0) {\cal U}_{\xi(T)}^\dagger,
\end{align}
by applying $e^{{\cal L}\ddt}\bullet$ repeatedly for $n = T/\delta t$ times to the initial state $\rho(0)$,  where we have defined
\be
{\cal U}_{\xi(T)} = \exp\left[ - i \hat H_{\rm S} T - i g\, \hat S \, \ddt \sum_{t=0}^{T-\ddt} \xi_t  \right],
\ee
which leads to the final state $\rho_{\xi}(T) = {\cal U}_{\xi(T)} \rho(0) {\cal U}_{\xi(T)}^\dagger$ of an unravelled trajectory. The function, 
\be\label{eq-pdfnoise}
\wp[\xi_0, \xi_{\ddt},..., \xi_{T-\ddt}] = \wp(\xi_0)\wp(\xi_{\ddt})...\wp(\xi_{T-\ddt}),
\ee
is the joint probability density function for a realization of the delta-correlated environmental noise's path, which makes the integration in Eq.~\eqref{eq-finalave} to be interpreted as over all possible noise's path realizations. Assuming that the target state $\rho_{\rm F}$ is a pure state, we can use the simplified definition of the fidelity in Eq.~\eqref{eq-fidel} and show that the fidelity to the average final state is
\begin{align}\label{eq-fidelave}
 \mathcal{F}[\rho(T),\rho_{\rm F}] &= {\rm Tr}[\rho(T)\rho_{\rm F}] \nonumber \\
 &= \int \! \prod_{t=0}^{T-\ddt} \dd \xi_t \,  \wp[\xi_0,..., \xi_{T-\ddt}] \, \mathcal{F}[\rho_{\xi}(T),\rho_{\rm F}]  \equiv {\bar {\cal F}}.
\end{align}
That is, the fidelity from the target state to the average final state is simply an average of fidelities from the target state to all possible final states $\rho_\xi(T)$ generated from the unravelled trajectories. The average is over all possible noise trajectories with the appropriate probability weights given by $\wp[\xi_0,..., \xi_{T-\ddt}]$ in Eq.~\eqref{eq-pdfnoise}.

From Eq.~\eqref{eq-fidelave}, we can see that, using the mean path approach, the objective for controlling the state preparation is to maximise the fidelity to all possible final states, $\rho_{\xi}(T)$, \textit{on average}. Note that an average value of a random variable does not necessarily reflect whether such value is attainable with a high probability. 
Therefore, it could be interesting to explore an alternative 
objective function that ensures the majority of the final states would concentrate around the target state. 
We then introduce a new objective function based on a probability density function defined for the unravelled quantum state trajectories of the open quantum system as in Eq.~\eqref{eq-unravelk1}. We will see in the next section that maximising the probability function can be conveniently achieved with the least action principle of a stochastic path integral.

\section{Stochastic path integral for state preparation and success rates}\label{sec3}
\subsection{Stochastic path integral formalism}

Stochastic path integral (SPI) is a mathematical technique used in representing probability density functions (PDFs) of stochastic processes that are continuous in time~\cite{BookSchulman,BookKleinert,WebErw2017}. It is typically constructed for Markovian stochastic processes, where a joint probability density function of a time-continuous process can be approximated as a product of probability functions of infinitesimal-time-discretized states. The SPI can then be extremised, similar to the action principle of the Feynman path integral \cite{BookFeyHib}, to obtain variational solutions. Such solutions can be interpreted as ``optimal" paths in the sense that they maximise some kinds of probability functions of the stochastic processes~\cite{BookKamenev}. Among various proposals in the literature, here we use the SPI developed for a continuous quantum measurement in \cite{Chantasri2013,chantasri2015stochastic,chantasri2016,LewCha17,KarLew2022}. In particular, we modify the technique to suite the quantum control problem for unravelled stochastic trajectories.

In the original work \cite{Chantasri2013,chantasri2015stochastic}, the SPI is constructed from the joint probability density functions of continuous measurement readouts and the corresponding quantum state trajectories. We follow similar mathematical reasonings, but replacing the continuous readouts with environmental noises that affect the quantum system's states and adding continuous controls. Let us first discretise a time range $[0,T]$ into a set $\{0, \ddt, 2\ddt,...,n\ddt \}$ where $n\ddt = T$. Given the environmental noises ${\bm \xi}_t$ (here, we use bold letters to indicate vectors for general multi-dimensional variables), where $t\in [0,T]$, we define a set of discretized noises for all times as, $\both{\bm \xi} \equiv \{ {\bm \xi}_t : t \in \{0,\ddt, 2\ddt... , T-\ddt\}\}$, a time-continuous control, $\both{\bm u} \equiv \{ {\bm u}_t : t \in \{0,\ddt, 2\ddt... , T-\ddt\}\}$, and their corresponding quantum states, $\bothp{\bm q} \equiv \{ {\bm q}_t : t \in \{0,\ddt, 2\ddt, ... , T\}\}$. Noting that the arrowheads indicate different time ranges. Assuming that the stochastic process is Markovian, the system's state at any time $t+\ddt$ can be written as a function of the state, the noise, and the control, at the intermediate past step $t$. That is, we have
\be\label{eq-statemap}
\boldsymbol{q}_{t+\ddt} =\boldsymbol{\mathcal{E}}(\boldsymbol{q}_t,{\bm \xi}_t, {\bm u}_t),
\ee
for an evolution function $\boldsymbol{\mathcal{E}}(\bullet)$. This is a control-extended version of the quantum measurement process, where $\boldsymbol{q}_{t+\ddt} =\boldsymbol{\mathcal{E}}(\boldsymbol{q}_t,{\bm \xi}_t)$, used in the original work~\cite{chantasri2015stochastic}. \red If the information of noise can be partially obtained, such information can be included in this evolution function.\blk

Following Refs.~\cite{Chantasri2013,chantasri2015stochastic}, we start by constructing a joint probability density function (PDF) of the time-discretized quantum trajectories and  their environmental noises, given the control function and an initial state $\boldsymbol{q}_0$,
\begin{equation} \label{PDF}
 \wp(\, \bothp{\bm q},\, \both{\bm \xi}\, |\,  \both{\bm u}, \boldsymbol{q}_0)
= \mathcal{B}\prod_{t=0}^{T-\ddt}
\wp(\boldsymbol{q}_{t+\ddt}|\boldsymbol{q}_t, {\bm \xi}_t, {\bm u}_t) \wp({\bm \xi}_t|\bb{q}_t),
\end{equation}
using Bayes' rules. The right-hand side of Eq.~\eqref{PDF} is written as a discrete-time product of PDFs of noises, ${\bm \xi}_t$, which only depends on the concurrent states, ${\bm q}_t$, multiplied by PDFs of intermediate-future states, ${\bm q}_{t+\ddt}$, given the states, ${\bm q}_t$, the noises, ${\bm \xi}_t$, and the controls, ${\bm u}_t$. According to Eq.~\eqref{eq-statemap}, we can write $\wp(\boldsymbol{q}_{t+\ddt}|\boldsymbol{q}_t,{\bm\xi}_t, {\bm u}_t)\equiv \delta(\boldsymbol{q}_{t+\ddt}-\boldsymbol{\mathcal{E}}(\boldsymbol{q}_t,{\bm \xi}_t, {\bm u}_t))$ as a delta-function probability density function. Moreover, we are interested in the case of quantum control for the state preparation, where the initial and final states of the quantum system are fixed at ${\bm q}_0 = {\bm q}_I$ and ${\bm q}_T = {\bm q}_F$, respectively. The boundary term in Eq.~\eqref{PDF} is then given by the delta functions $\mathcal{B} = \delta({\bm q}_0 - {\bm q}_I)\delta({\bm q}_T - {\bm q}_F)$. 

The joint PDF in Eq.~\eqref{PDF} can be transformed into a path integral by rewriting all the delta functions in their Fourier representation, i.e., $\delta(q)=(1/2\pi i)\int_{-i \infty}^{i \infty} \!\dd p \, \exp(-p q)$, where $p$ is a conjugate variable, and recasting the rest of the terms in exponential forms. We can apply this to all components of the vector $\bm q$. As a result, by introducing a set of conjugate variables $\bothp{\bm p} = \{ {\bm p}_t : t \in \{-\ddt, 0,\ddt, ... , T\}\}$, we obtain the SPI
\be\label{eq-spi}
\wp(\, \bothp{\bm q},\, \both{\bm \xi}\, | \, \both{\bm u}, \boldsymbol{q}_0)=\mathcal{N} \!\! \int \!\! \prod_{t = -\ddt}^{T} \!\! \dd{\bm p}_t \, \exp\mathcal{S}[ \, \bothp{\bm p},\,   \bothp{\bm q}, \,  \both{\bm \xi}, \, \both{\bm u}\, ],
\ee
where the integral is over all possible paths of ${\bm p}_t$ and the \textit{action}, ${\cal S}$, is given by,
\begin{align}\label{action}
\mathcal{S}[\,  \bothp{\bm p},\,   \bothp{\bm q}, \,  \both{\bm \xi}, \, \both{\bm u}\, ] \equiv &  - {\bm p}_{-\ddt}\cdot({\bm q}_0 - {\bm q}_I)  - {\bm p}_{T}\cdot({\bm q}_T - {\bm q}_F) \nonumber \\ 
&+\sum_{t=0}^{T-\ddt}\{-\boldsymbol{p}_t\cdot[\boldsymbol{q}_{t+\ddt}-\boldsymbol{\mathcal{E}}(\boldsymbol{q}_t,{\bm \xi}_t, {\bm u}_t)]+\ln \wp({\bm \xi}_t|{\bm q}_t)\} ,
\end{align}
where any constant factors left are absorbed into the factor ${\cal N}$ in Eq.~\eqref{eq-spi}.

Given the stochastic path integral, one can use the variational principle and solve for the extrema of the action in Eq.~(\ref{action}). This can be done by taking the variation of the action over all the variables and setting it to zero. That is
\begin{equation}\label{eq-variation}
\bb{\n}_{{\bm p}_t}\mathcal{S} =0,\,\, \bb{\n}_{{\bm q}_t}\mathcal{S} =0,\,\,
\bb{\n}_{{\bm \xi}_t}\mathcal{S} =0,\,\, \mbox{and  }\,\, \bb{\n}_{{\bm u}_t}\mathcal{S} =0,
\end{equation}
where $\bb{\n}_{{\bm x}_t}$ denotes the gradient with respect to a vector ${\bm x}_t$, which respectively leads to a set of the difference equations:
\begin{subequations}\label{eq-diffeq}
\begin{eqnarray}
	\bb{q}_{t+\ddt}&=& \bb{\mathcal{E}}(\bb{q}_t,{\bm \xi}_t, {\bm u}_t),\\
	\bb{p}_{t-\ddt}&=&\bb{\n}_{\bb q_t}\left[ \,  {\bm p}_t \cdot \bb{\mathcal{E}}(\bb{q}_t,{\bm \xi}_t, {\bm u}_t)+\ln \wp({\bm \xi}_t|{\bm q}_t)\right] ,\\	
	0&=&\bb{\n}_{{\bm \xi}_t}\left[ \,  {\bm p}_t \cdot \bb{\mathcal{E}}(\bb{q}_t,{\bm \xi}_t, {\bm u}_t)+\ln \wp({\bm \xi}_t|{\bm q}_t)\right],\\
	0&=&\bb{\n}_{{\bm u}_t}\left[ \,  {\bm p}_t \cdot \bb{\mathcal{E}}(\bb{q}_t,{\bm \xi}_t, {\bm u}_t)+\ln \wp({\bm \xi}_t|{\bm q}_t)\right].
\end{eqnarray}
\end{subequations}
\noindent In most cases, these difference equations approximate differential equations that can be solved numerically, or in some cases even analytically. The differential equations can be obtained by taking the limit $\ddt \rightarrow \dd t$. The solutions obtained are ``extremal" functions, denoted by ${\bm q}^*_t$, ${\bm \xi}^*_t$, and ${\bm u}^*_t$, which can be respectively interpreted as ``optimal" state trajectory, noise path, and control function. We note that one can also treat the action in Eq.~\eqref{action} as a Lagrange function for the method of Lagrange multipliers, where the function to be mimimised or maximised is $\sum_{t=0}^{T-\ddt}\ln \wp({\bm \xi}_t|{\bm q}_t)$, i.e., the log of the probability density function of a noise path, under the constraints of the quantum state evolution and the fixed boundary states~\cite{LewCha17}.

Since Eq.~\eqref{eq-variation} is the first order condition, the extremal solution could represent a local minimum, a local maximum, or a saddle point. In this work, however, we are interested in the solution that maximises the action, $\mathcal{S}[\,  \bothp{\bm p},\,   \bothp{\bm q}, \,  \both{\bm \xi}, \, \both{\bm u}\, ]$, where the extremal solution can be regarded as the ``most-likely path (MLP)" or a quantum path (trajectory) associated with the most-likely noise.  Therefore, the SPI and its variational solutions in Eqs.~\eqref{eq-diffeq} provide us with a convenient approach to solve for time-continuous controls ${\bm u}^*_t$ subjected to the most-likely noise ${\bm \xi}^*_t$, given the fixed initial and final states ${\bm q}_I$ and ${\bm q}_F$, respectively.

\subsection{Success rate as a metric for quantum control}

As we mentioned in Section~\ref{sec-avefid}, the Lindblad (mean path) approach does maximise the average fidelity over all possible noise realizations, as shown in Eq.~\eqref{eq-fidelave}. Also, as mentioned, maximising the average fidelity does not necessarily guarantee a high chance in reaching the target state. Therefore, we instead propose the use of the most-likely path, a solution of Eqs.~\eqref{eq-diffeq}, which maximises the noise-path probability density function given the fixed boundary states. The solution includes a control function $\bm u^*_t$ that could potentially maximise the chance to reach the target state. 

In order to benchmark this new approach against the conventional mean-path one, e.g., in a numerical simulation, we  
therefore need to define a new quality metric that is closely related to the likelihood of noise. We propose a measure called \textit{success rate}, defined as a percentage ratio of unravelled noisy trajectories (as results of a particular control function) whose final states are close to the target state within some \textit{infidelity tolerance} (IT), denoted by $\delta$. That is, given a choice of quantum control, the success rate is computed from
\be\label{eq-SR}
s (\delta) := \frac{N_\delta[ {\cal F}({\bm q}_T, {\bm q}_F) \ge (1- \delta)]}{N_{\rm tot}} \, \%,
\ee
where $N_\delta[\cal C]$ is the number of noise realizations that satisfy the condition $\cal C$, and $N_{\rm tot}$ is the total number of noise realizations considered. The success rate thus reflects how much the ensemble of noisy final states concentrates around the desired target state. In other words, a high success rate indicates a high chance that the actual (but unknown) final state could reach the target state.

It is important to note that neither the MP nor MLP approach is mathematically constructed to directly maximise the heuristic success rate defined here. So, we cannot expect the MLP control to always give higher success rates than the MP one. However, since the MLP approach does maximise the noise-path probability density given the fixed target state, we expect that most of possible noise trajectories will concentrate around the most-likely path and then contribute to the high chance of reaching the target state~\cite{chantasri2015stochastic}. 
We will show later in Section~\ref{sec5} that the MLP control in many situations yields higher success rates (hence, a higher quality ensemble of the final states) than the MP control that maximises the standard average fidelity metric (Eq.~\eqref{eq-fidel}). We also note that in order to come up with a quantum control strategy that directly maximises the success rate of Eq.~\eqref{eq-SR}, one would need to numerically solve for the full Fokker-Planck equation for the noisy quantum states to obtain the final state distribution, then integrate the distribution over the infidelity tolerance, and then search over all possible control functions to maximise the infidelity integral. Such method could be done in theory, but it will take an incredible amount of computational resource for an insignificant improvement that the MLP or even the MP approaches can already provide, which is beyond the scope of our paper.

\section{Optimal Rabi Drive for Qubit State Preparation}\label{sec4}

In this section, as a proof of principle, we apply the standard mean-path (MP) approach of Section~\ref{sec-mp-app} and our proposed most-likely path (MLP) of Section~\ref{sec3} to a two-level system coupled to a noisy environment. For the MP approach, we obtain a Lindblad equation that can be used in the numerical search for optimal controls. For the MLP approach, we obtain a set of ODEs which can be solved analytically for the optimal control. 

\subsection{Qubit model with a pure-dephasing noise and a controlled Rabi drive}\label{sec4-1}
In order to obtain the unravelled system-environment Hamiltonian as in Eq.~\eqref{eq-hamtot}, we need to consider a noisy environment that interacts with a Hermitian observable $\hat S$ of the system. One of the most typical noises in qubit experiments that fits this criterion is the pure dephasing noise~\cite{BookGardiner}. It is the noise that causes the energy fluctuations in qubits and leads to $\hat S \propto \hat \sigma_z$. 
We assume \red that, in the rotating frame with the fast qubit's bare frequency and the bath's bare Hamiltonian, the qubit has a small leftover \blk energy gap $\epsilon$ 
and a Rabi oscillation with a controllable time-dependent frequency $\Omega(t)$. The controlled Rabi frequency is motivated by the typical Rabi control via on-resonance driving fields. Therefore, we obtain the qubit's Hamiltonian as $\hat{H}_{\rm S} = (\epsilon/2)\hat\sigma_z-(\Omega(t)/2)\hat\sigma_x$, where, in the Bloch sphere representation, the $\epsilon$ and $\Omega(t)$ indicate the angular speeds of qubit's rotations around the Bloch sphere's $z$- and $x$-axes, respectively.
Combining the qubit's unitary dynamics and the effect from a pure dephasing noise, we have the Lindblad evolution, $\rho(t+\delta t) = \exp({\cal L} \delta t)\rho(t)$, similar to Eq.~(\ref{eq-decoh}), where\blk
\be\label{eq-qubitLB}
{\cal L} \rho(t)=-i  \left[\frac{\epsilon}{2}\hat\sigma_z-\frac{\Omega(t)}{2}\hat\sigma_x ,\rho(t) \right]+\frac{\gamma}{2} \, \mathcal{D}[\hat \sigma_z ]\rho(t),
\ee
and $\gamma$ indicates the qubit's dephasing rate from coupling to the noisy environment.

As we discussed in Section \ref{sec-unravel}, the Lindblad master equation describing an averaged evolution (mean path) can be unravelled into stochastic evolutions. Let us use the same variables as before: the environmental noise variable $\xi(t)$ \red (which can be considered as one of the eigenvalues of the bath's observable $\hat B(t)$ as described in Section~\ref{sec-unravel}) \blk and the coupling rate $g$. The Hamiltonian for the unravelled stochastic evolution, similar to Eq.~\eqref{eq-stoHamil}, is given by
\begin{align}\label{H}
	\hat{H}_{\xi}(t)\,&= \hat H_{\rm S} + g\, \xi(t)\, \hat S \nonumber \\
	&= \frac{\epsilon}{2}\hat\sigma_z-\frac{\Omega(t)}{2}\hat\sigma_x+g\, \xi(t)\, \hat \sigma_z .
\end{align}
For the pure dephasing noise, one can generally assume that $\xi(t)$ is a Gaussian white noise with the zero mean, $\langle \xi(t) \rangle = 0$, and the delta-function correlation, $\langle \xi(t)\xi(t') \rangle = \kappa\delta(t-t')$. The Gaussian white noise approximation is the simplest yet ubiquitous noise representation observed in experiments.

This unravelled stochastic Hamiltonian of Eq.~\eqref{H} can be used in deriving the Lindblad evolution and constructing the state mapping Eq.~\eqref{eq-statemap} for the most-likely path approach. Let us first switch to the Bloch sphere representation for qubit's states and use the time-discrete notation. For a qubit density matrix $\rho$, we define a Bloch vector from 
\be
{\bm q} = \{ x,y,z\} \equiv \{ {\rm Tr}[\hat \sigma_x \rho], {\rm Tr}[\hat \sigma_y \rho], {\rm Tr}[\hat \sigma_z \rho] \}.
\ee
For the time-discrete variables, we have a set of the states $ \{ {\bm q}_0,..., {\bm q}_t, ..., {\bm q}_T\}$, the noises, $\both{\xi} = \{ \xi_0,..., \xi_t,..., \xi_{T-\ddt}\}$, and the control variables, $\both{u}= \{\Omega_0,..., \Omega_t,..., \Omega_{T-\ddt}\}$. The time-discrete version of the unitary unravelled dynamics is given by 
\be\label{eq-disumap}
\rho_{t+\ddt}  = \exp\left[-i  \hat{H}_{\xi_t} \ddt\right] \rho_t \exp\left[+i \hat{H}_{\xi_t} \ddt \right].
\ee
where the unravelled Hamiltonian $\hat H_{\xi}(t)$ in Eq.~\eqref{H} becomes
\be
\hat{H}_{\xi_t} \equiv \frac{\epsilon}{2}\hat \sigma_z - \frac{\Omega_t}{2} \hat\sigma_x+g \, \xi_t\, \hat \sigma_z.
\ee
We can then write Eq.~\eqref{eq-disumap} in the Bloch sphere coordinate and expand the right-hand side to the second order in $\delta t$ to get
\begin{eqnarray}\label{SDE1}
		x_{t+\delta t}&=&x_t-\delta ty_t\tilde{\epsilon}_t-\frac{\delta t^2}{2}(z_t\Omega_t\tilde{\epsilon}_t+x_t\tilde{\epsilon}_t^2)+\mathcal O(\delta t^3), \nonumber \\
		y_{t+\delta t}&=&y_t+\delta t(x_t\tilde{\epsilon}_t +z_t\Omega_t)-\frac{\delta t^2}{2}(y_t\Omega^2_t+y_t\tilde{\epsilon}_t^2)+\mathcal O(\delta t^3), \nonumber \\
		z_{t+\delta t}&=&z_t-\delta ty_t\Omega_t-\frac{\delta t^2}{2}(x_t\Omega_t\tilde{\epsilon}_t+z_t\Omega_t^2)+\mathcal O(\delta t^3),
\end{eqnarray}
where we have defined $\tilde \epsilon_t = \epsilon + 2 g \xi_t$ for convenience. From here on, we will use Eq.~\eqref{SDE1} as the time-infinitesimal state map, ${\bm q}_{t+\ddt} = \boldsymbol{\mathcal{E}}(\boldsymbol{q}_t,{ \xi}_t, \Omega_t)$ in Eq.~\eqref{eq-statemap}, for our qubit example. The reason we have expanded the state update equation to second order in $\ddt$ is so that we can derive the average dynamics, via the It\^o's prescription: $\xi_t^2 \ddt^2 \approx \kappa \ddt$ \cite{BookGardiner2}, which needs terms expanded to ${\cal O}(\ddt^2)$ (see Section~\ref{sec-exMP}) and that we can also derive the MLP equations via Eq.~\eqref{eq-diffeq}, which needs terms expanded to ${\cal O}(\ddt)$ (see Section~\ref{sec-exMLP}).


\subsection{Mean path (MP) approach}\label{sec-exMP}
Let us first use the MP approach to solve for the optimal Rabi drive that can prepare a final state at time $T$, ${\bm q}_T = \{ x_T, y_T, z_T\}$, that is close in fidelity to the target state ${\bm q}_{\rm F} =\{ x_{\rm F}, y_{\rm F}, z_{\rm F}\}$ given that the system is initialized at ${\bm q}_{\rm I} = \{ x_{\rm I}, y_{\rm I}, z_{\rm I}\}$. The mean-path dynamics for the qubit's example is given by the Lindblad master equation in Eq.~\eqref{eq-qubitLB}. However, we can also show that the master equation can be obtained from averaging the unravelled dynamics in Eq.~\eqref{SDE1}. To do that, we first derive the It\^o stochastic differential equation by taking $\xi_t^2 \delta t^2\approx \kappa\delta t $ in Eq.~\eqref{SDE1} and then take an average over the Gaussian-white noise ensemble (equivalent to replacing $\xi_t =0$ in the It\^o equation). We obtain
\begin{subequations}\label{eq-ito}
\begin{eqnarray}
		x_{t+\delta t}&=&x_t-\delta t(y_t\epsilon +2g^2\kappa x_t),\\
		y_{t+\delta t}&=&y_t+\delta t(x_t\epsilon +z_t\Omega_t -2g^2\kappa y_t),\\
		z_{t+\delta t}&=&z_t-\delta ty_t\Omega_t,
\end{eqnarray}
\end{subequations}
which, after taking the time-continuum limit $\delta t\rightarrow{\dt}$, gives back the Lindblad master equation Eq.~\eqref{eq-qubitLB} with the dephasing rate $\gamma=2g^2\kappa$. 


To search for the optimal control (optimal Rabi drive) using the MP approach, the average dynamics, Eqs.~\eqref{eq-ito}, should be used as a representative of the quantum state dynamics. However, because of the dephasing effect, the qubit state at any time $T$ can never be a pure state and thus Eqs.~\eqref{eq-ito} cannot have a solution that exactly satisfies ${\bm q}_T = {\bm q}_{\rm F}$ for the pure target state. The best one could do is to search for a Rabi control function, $\Omega_t = \Omega^{\rm MP}_{\rm op}$, that maximises the qubit fidelity
\be\label{eq-avefid-qubit}
{\cal F}_{q}[{\bm q}_T,{\bm q}_{\rm F}]  \equiv {\cal F}[\rho(T),\rho_{\rm F}] =\frac{1}{2}(1+x_T x_{\rm F}+y_T y_{\rm F}+z_T z_{\rm F}) .
\ee
Since an analytical solution is not available, one resorts to numerical methods. The search of an infinite functional space of the Rabi control function can be simplified by dividing the control into $m$ time intervals, $\{ \Omega_j \} = \{\Omega_1, \Omega_2, \dots, \Omega_m\}$, assuming $\Omega_j$ be constant in each of the intervals, then one can use the quantum control algorithms, e.g., GRAPE and CRAB algorithms to search for $\Omega^{\rm MP}_{\rm op}$ that maximises the fidelity in Eq.~(\ref{eq-avefid-qubit}). 

\subsection{Most-likely path (MLP) approach}\label{sec-exMLP}
An alternative method is to use the MLP approach, as described in Section~\ref{sec3}, to search for an optimal Rabi control. Given the parameters defined in Section~\ref{sec4-1}, we have the state update, $\boldsymbol{q}_{t+\ddt}=\boldsymbol{\mathcal{E}}[\boldsymbol{q}_{t}, \xi_t, \Omega_t]$, given by Eq.~\eqref{SDE1}. For the probability density function of the delta-correlated noise we introduced in Eq.~\eqref{H}, we have
\be\label{eq-probxi}
\wp(\xi_t | {\bm q}_t) = \wp(\xi_t) = \sqrt{\frac{\ddt}{2\pi \kappa}}\exp\left( - \xi_t^2 \ddt/2 \kappa\right),
\ee
because $\xi_t$ is assumed a Gaussian white noise, which is also independent of the qubit state. For the boundary term $\mathcal{B}= \delta(\boldsymbol{q}_0-\boldsymbol{q}_{\rm I})\delta(\boldsymbol{q}_{T}-\boldsymbol{q}_{\rm F})$, the two fixed boundary states, $\boldsymbol{q}_{\rm I}$ and $\boldsymbol{q}_{\rm F}$, are naturally the initial state  and the desired target state of the state-preparation problem, respectively. Following section~\ref{sec3}, we construct the stochastic path integral from Eq.~(\ref{eq-spi}), with the action in Eq.~(\ref{action}), which gives
\begin{align}
\label{eq-action-mlp}
\mathcal{S}=& - {\bm p}_{-\ddt}\cdot({\bm q}_0 - {\bm q}_{\rm I})  - {\bm p}_{T}\cdot({\bm q}_T - {\bm q}_{\rm F}) \\
&+\sum_{t=0}^{T-\ddt}\{-\boldsymbol{p}_t\cdot[\boldsymbol{q}_{t+\ddt}-\boldsymbol{\mathcal{E}}(\boldsymbol{q}_t,{\xi}_t, {\Omega}_t)]-\xi^2_t \delta t/2 \kappa \}\nonumber,
\end{align}
where we have introduced ${\bm p}_t \equiv \{ p_{x,t}, p_{y,t}, p_{z,t}\}$ as the conjugate variables of the Bloch sphere vector ${\bm q}_t = \{ x_t, y_t, z_t\}$. The last term in the action is from the log probability, $\ln \wp(\xi_t) = - \xi_t^2 \ddt/2 \kappa + \ln(\sqrt{\ddt/2\pi \kappa})$, in Eq.~\eqref{eq-probxi}, where the second (constant) term can be absorbed into the path integral measure. In order to solve for the most-likely path, we use the variational principle, extremising the action over all variables at all time steps, i.e.,
\begin{equation}\label{eq-extremize}
\bb{\n}_{{\bm p}_t} {\cal S} = 0,\,\, \bb{\n}_{{\bm q}_t} {\cal S} = 0,\,\,
\partial_{\xi_t}{\cal S}=0,\,\,\mbox{and  }\,\,\partial_{\Omega_t}{\cal S}=0,
\end{equation}
similar to Eq.~\eqref{eq-variation}. The first two conditions lead to explicit equations of motion of the most-likely path described by six ordinary difference equations:
\begin{subequations}\label{eq-6ODEs}
\begin{eqnarray}
x_{t+\delta t}&=&x_t-\delta t y_t\tilde{\epsilon}_t,\\
y_{t+\delta t}&=&y_t+\delta t(x_t\tilde{\epsilon}_t +z_t\Omega_t),\\
z_{t+\delta t}&=&z_t-\delta t y_t\Omega_t,\\
p_{x,t+\delta t}&=&p_{x,t}-\delta t p_{y,t}\tilde{\epsilon}_t,\\
p_{y,{t+\delta t}}&=&p_{y,t}+\delta t(p_{x,t}\tilde{\epsilon}_t +p_{z,t}\Omega_t),\\
p_{z,{t+\delta t}}&=&p_{z,t}-\delta t p_{y,t}\Omega_t,
\end{eqnarray}
\end{subequations}
keeping terms to only first order in $\delta t$, with $\tilde \epsilon_t = \epsilon + 2 g \xi_t$ as before. The last two constraints of Eq.~(\ref{eq-extremize}) give:
\begin{subequations}\label{eq-constraint}
\begin{eqnarray}
 \xi_t&=&  -2g \kappa p_{x,t}y_t+2g \kappa p_{y,t}x_t\label{eq-con1},\\
p_{y,t}z_t&=&p_{z,t}y_t\label{eq-con2}. 
\end{eqnarray}
\end{subequations}
These equations can be solved in the time-continuum limit ($\ddt \rightarrow \dt$), where the first six difference equations become six ordinary differential equations (ODEs) and Eqs.~\eqref{eq-con1} and \eqref{eq-con2} are the two constraints. As mentioned earlier, this technique is similar to the Lagrange multiplier method, where the solutions of these equations give a quantum state path, $x_t^*$, $y_t^*$, $z_t^*$, a noise path, $\xi_t^*$, and a control function, $\Omega_t^*$, that maximise the log-likelihood of the noise path, i.e., maximising $\ln \prod_{t=0}^{T-\ddt} \wp(\xi_t)$. 

In an attempt to solve for optimal solutions analytically, we find that Eq.~(\ref{eq-con2}) is a canonical relation between conjugated variables and the components of the Bloch sphere vector. Taking time derivatives on both sides of Eq.~(\ref{eq-con2}) and substituting the relations in Eqs.~\eqref{eq-6ODEs} lead to another relation
\begin{eqnarray}		
p_{x,t} \, z_t=p_{z,t} \, x_t.\label{c2}
\end{eqnarray}
By substituting the above in Eqs.~\eqref{eq-con2} and \eqref{eq-con1}, we then get $\xi_t^*=0$. This seems to be an unsurprising result, considering that the noise is a zero-mean Gaussian white noise. However, it is important to note that the solution $\xi_t^* = 0$ is the \textit{most-likely noise path}, which gives the dynamics following Eqs.~\eqref{eq-6ODEs}, and {\it not the average dynamics} of Eqs.~\eqref{eq-ito}.

Let us consider a single-pulse case ($\Omega_t = \Omega$), which admits full analytical solutions for the MLP control. By imposing the zero-noise path condition, $\xi_t^* = 0$, all the conjugate variables vanish, reducing the six ODEs in Eqs.~\eqref{eq-6ODEs} to only three ODEs for the coordinate variables that can be solved analytically to yield the most-likely path
\begin{subequations}\label{eq-solution}
\begin{eqnarray}
		x_t^*&=&-\frac{\epsilon\, \Omega}{\omega^2}\cos(\omega t)+\frac{\epsilon\, \Omega}{\omega^2},\\
		y_t^*&=&-\frac{\Omega}{\omega}\sin(\omega t),\\
		z_t^*&=&-\frac{\Omega^2}{\omega^2}\cos(\omega t)-\frac{\epsilon^2}{\omega^2},
\end{eqnarray}
\end{subequations}
where $\omega\equiv\sqrt{\Omega^2+\epsilon^2}$. Solving the above equations for the fixed boundary conditions: $\{ x^*_0 ,y^*_0, z^*_0\} = \{ x_{\rm I}, y_{\rm I}, z_{\rm I}\}$ and $\{ x^*_T ,y^*_T, z^*_T\} = \{ x_{\rm F}, y_{\rm F}, z_{\rm F}\}$ gives unique solutions for the Rabi drive and the total time of the evolution, simultaneously. We obtain the optimal values for $\Omega$ and $T$ as
\begin{subequations}\label{RabiTimeMLP}
\begin{eqnarray}
    \label{RabiMLP} \Omega_{\rm op}^{\rm MLP} &=& \left(\frac{z_{\rm F}-z_{\rm I}}{x_{\rm F}-x_{\rm I}}\right)\epsilon,\\ 
    \label{TimeMLP}
    T_{\rm op}^{{\rm  MLP}}&=&\frac{1}{\omega}\, {\arccos}\left[1-\frac{\omega^2}{\epsilon^2}\frac{x_{\rm{ F}}^2}{(z_{\rm{ I}}^2-z_{\rm{ I}}z_{\rm{ F}})}\right],
\end{eqnarray}
\end{subequations}
which are functions of the boundary states. We note that, for convenience, we show the optimal time Eq.~\eqref{TimeMLP} for $\{x_{\rm I} , y_{\rm I}, z_{\rm I}\} = \{ 0,0,z_I\}$, where its full formula is lengthy and shown in Appendix~\ref{app-diff}. It is important to note that, in contrast to the MP approach whose analytical solutions are not known, the final state of this most-likely path is \textit{exactly at the target state} ${\bm q}_{\rm F}$. Moreover, the qubit states along the most-likely evolution are all pure states, representing a possible qubit unravelled trajectory with the zero-noise trajectory.

\begin{figure}[t]
\centering
\includegraphics[width=12cm]{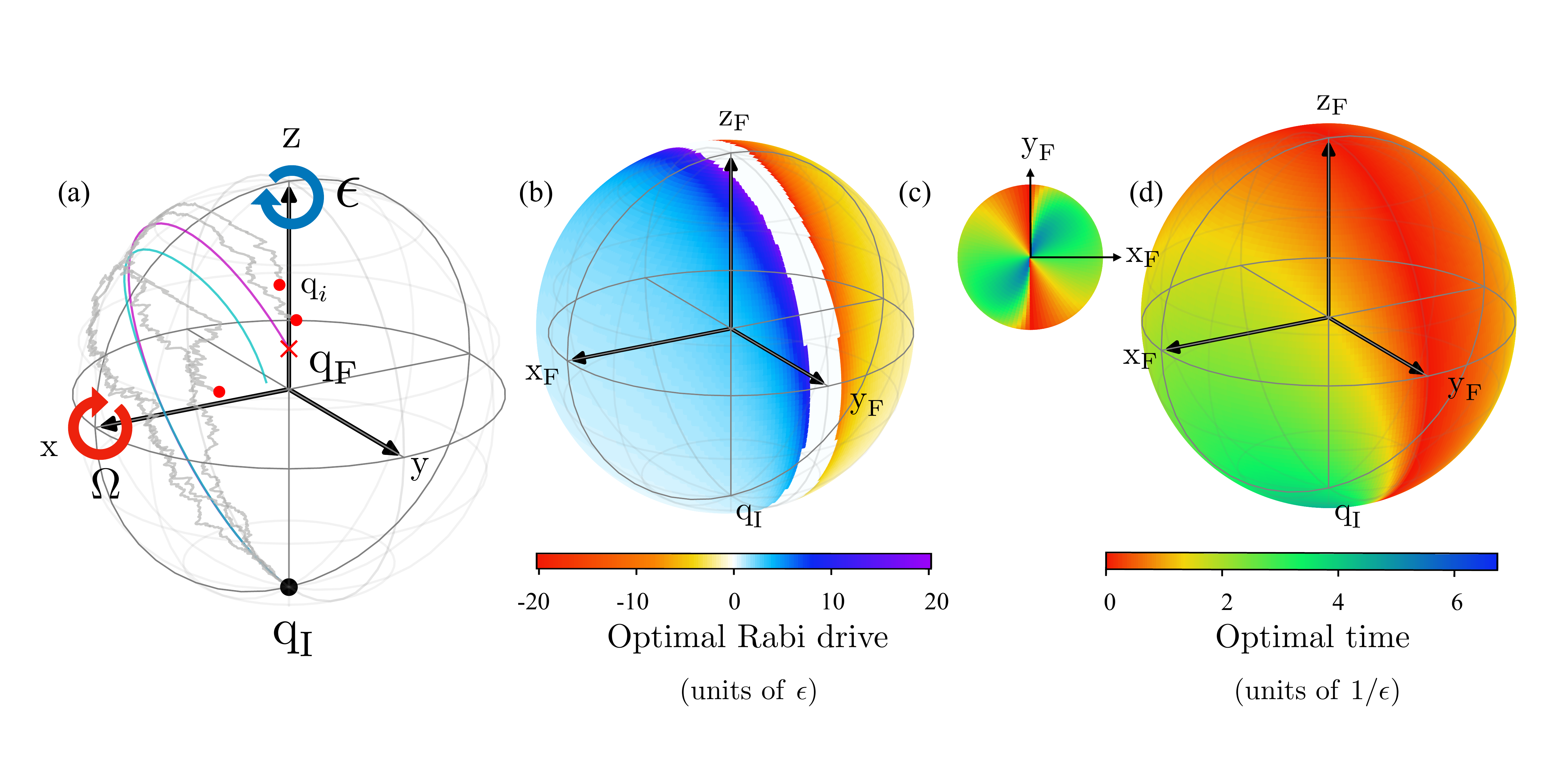}
\caption{The Bloch sphere plots describing our qubit model with the pure dephasing noise and the solutions of optimal Rabi drive and time in Eqs.~\eqref{RabiTimeMLP}. (a) Our qubit-state preparation model with the initial ground state ${\bm q}_{\rm I}$ (black dotted) and a target state (red cross) chosen to be ${\bm q}_{\rm F} = \{\sin(2\pi/3)/2,0.75,0.5\}$. The blue and red circled arrows indicate the rotational dynamics from $\epsilon$ and $\Omega$ terms in the qubit Hamiltonian. 
The optimal mean path (cyan curve), a solution of Eqs.~(\ref{eq-ito}), and the most-likely path (magenta curve), a solution of Eqs.~(\ref{eq-solution})-(\ref{RabiTimeMLP}), are shown, along with a few stochastic (unravelled) trajectories (grey curves ending at red dots, showing their random final states). The MLP\red, as expected, \blk is the only path that exactly reaches the target state \red as it was modeled based on a single realization of noise instead of the dephasing effect\blk. (b) and (d) display the contour plots on the target-state Bloch sphere showing values of the optimal Rabi drive and the optimal time in Eq.~(\ref{RabiMLP}) and Eq.~\eqref{TimeMLP}, respectively. The transparent (noncoloured) band in (b) indicates that the optimal Rabi drive solution diverges ($|\Omega_{\rm op}^{\rm MLP}/\epsilon| \rightarrow \infty$) at the plane $x_{\rm F}=0$. (c) The projection of (d) on the $x$-$y$ plane for clarity.}
\label{fig-blochopt}
\end{figure}

The solutions of the optimal Rabi drive and the optimal time in Eqs.~\eqref{RabiTimeMLP} also possess a simple geometrical interpretation. From the qubit Hamiltonian, Eq.~\eqref{H}, when $\xi(t) = 0$ and $\Omega(t) = \Omega$, the qubit evolution becomes a simple unitary rotation around the axis $\hat \omega = \epsilon \hat z - \Omega \hat x$, with an angular speed of $\omega = \sqrt{\Omega^2+\epsilon^2}$ (see the magenta curve in  Figure~\ref{fig-blochopt}(a) for an example of the MLP). Since the value of $\epsilon$ is fixed in the problem, the Rabi drive has to be adjusted such that the axis of rotation can bring the initial to the target state. It turns out that there is a unique solution for the axis of rotation; then the optimal time can then be calculated by dividing the angular distance between the two states with the angular speed $\omega$. We present a full derivation and discussion of the geometrical interpretation of the most-likely path in Appendix~\ref{app-derive}.

Since the optimal solutions in Eqs.~\eqref{RabiTimeMLP} are only functions of the boundary states, we can deliberately choose the initial state (also for the rest of the paper) to be the ground state of the qubit ${\bm q}_{\rm I} = \{ 0 , 0, -1\}$ and plot the optimal solutions as a function of the target state.
We show in Figure~\ref{fig-blochopt}(b), (c), and (d) the contour plots of the optimal Rabi drive and the optimal time of Eqs.~\eqref{RabiTimeMLP} for different target states ${\bm q}_{\rm F} = \{ x_{\rm F}, y_{\rm F}, z_{\rm F}\}$ on the Bloch sphere.

The MLP optimal Rabi controls, shown in Figure \ref{fig-blochopt}(b), are positive or negative depending on the sign of $x_{\rm F}$. This is expected as the qubit has a fixed rotation effect from the $\epsilon$-term, which breaks the symmetry of $\pm x_{\rm F}$. Also, the Rabi magnitudes grow monotonically as $|x_{\rm F}|$ decreases from 1 to 0, and diverge at $x_{\rm F} = 0$. This indicates that, given the initial state at the bottom of the Bloch sphere, it is harder to reach any target state in the $x_{\rm F}=0$ plane, as the $\epsilon$-term will always move the state out of the plane. Since the Rabi drive, that can be implemented in experiments, e.g., superconducting qubits, is typically limited in strength, we set a Rabi cut-off at around $\Omega \sim 20 \epsilon$, resulting in the transparent (noncoloured) band around $|x_{\rm F}| < 0.1$. However, we note that those final states in the $x_{\rm F}=0$ plane are reachable, only if somehow one can switch off $\epsilon = 0$ that causes the rotation around the $z$-axis. Then, only a single Rabi drive around the $x$-axis would suffice to control the state to any targets on the $y$-$z$ plane.

In Figure~\ref{fig-blochopt}(c) and (d), we show contour plots for the optimal time. From the geometrical interpretation above, an optimal time is an angular distance between the two states ${\bm q}_I$ and ${\bm q}_{\rm F}$ divided by the angular speed (Rabi drive) associated with the target ${\bm q}_{\rm F}$. We will see in the next section that the values of the Rabi drive and the optimal time have a significant effect on the success rate of the MLP-control approach.


\section{Numerical analysis comparing MLP and MP optimal controls}\label{sec5}
In this section, we numerically investigate the qubit example introduced in the previous section aiming to compare the optimal control  from the MLP approach, $\Omega_{\rm op}^{\rm MLP}$, with the control found with the traditional MP approach, $\Omega_{\rm op}^{\rm MP}$. We numerically simulated unravelled stochastic qubit trajectories following Eq.~\eqref{SDE1}, where $\epsilon$ is chosen as a unit of our problems, the time step is $\delta t = 0.01\epsilon^{-1}$, and the trajectory ensemble size is $N_{\rm tot} = 10^{4}$.
We start by analysing qubit trajectories and distributions of the qubit's final states and then explore the success rate metric for a range of target states and different dephasing rates to obtain an overview of the performance of the controls. We also compare our proposed MLP control with the triple-pulse optimised controls via GRAPE and CRAB algorithms.



\subsection{Qubit state trajectories and final-state distributions}

\begin{figure*}[b]
\centering
\includegraphics[width=13cm]{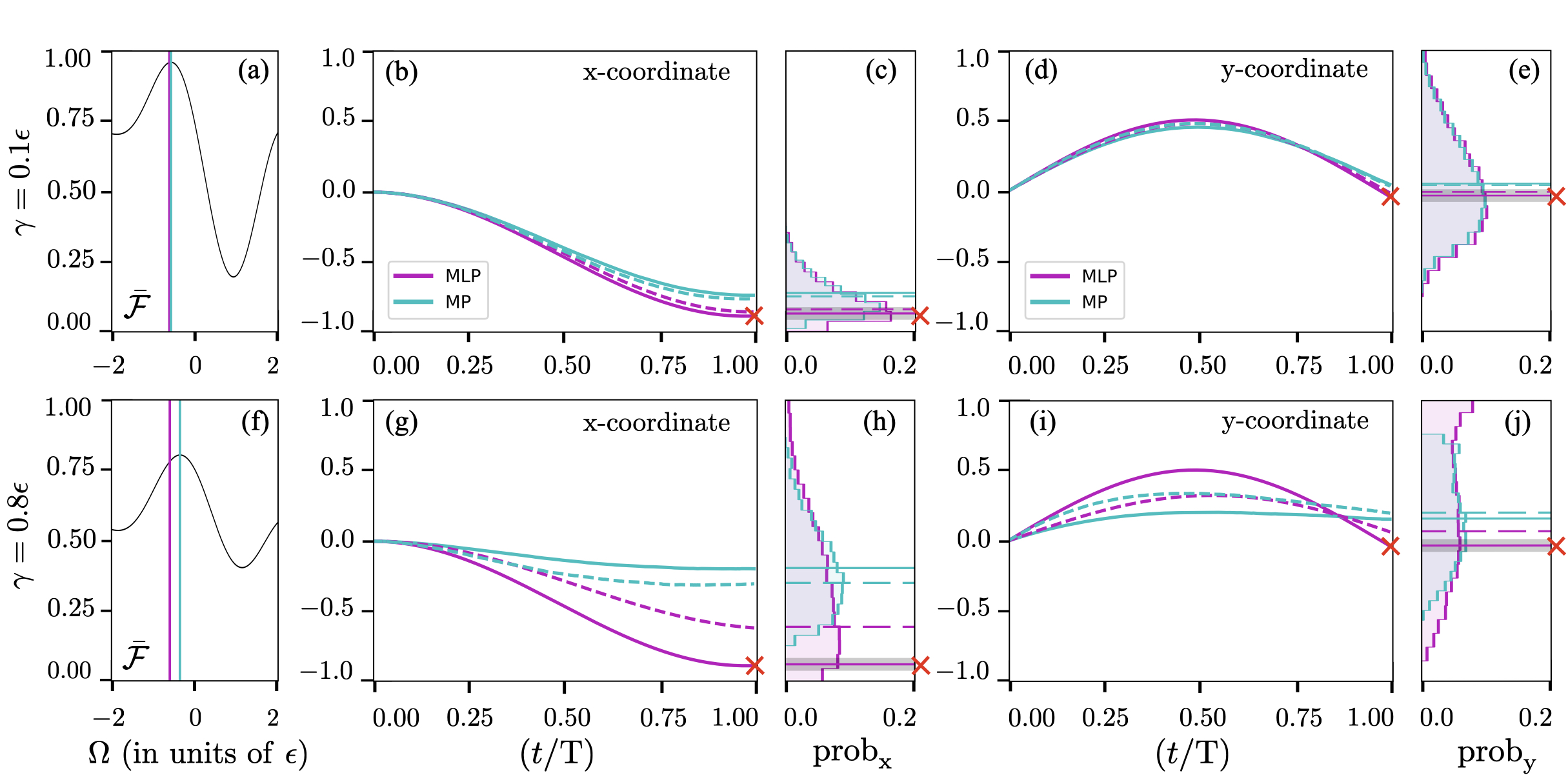}
\caption{ 
Examples of the MP and MLP Rabi controls, where the initial state is fixed at ${\bm q}_{\rm I}=\{0,0,-1\}$ and the target state is ${\bm q}_{\rm F} = \{-\sin(2\pi/3),0,-0.5\}$, for two values of the dephasing rate: $\gamma=0.1\epsilon$ (top row) and $\gamma=0.8\epsilon$ (bottom row). Colours (cyan and magenta) are chosen for the MP and MLP approaches, respectively. Panels (a) and (f) are plots of the average fidelity \red defined in Eq.~\eqref{eq-fidelave} (where each individual ${\cal F}$ follows Eq.~\eqref{eq-avefid-qubit} for the qubit fidelity), shown as a function of $\Omega$\blk. In (a), we found $\Omega_{\rm op}^{\rm MP} =  -0.55\epsilon$ and $\Omega_{\rm op}^{\rm MLP} = -0.57\epsilon$, giving $\bar{\mathcal{F}}_{\rm MP} =  0.9506$ \red larger than \blk $\bar{\mathcal{F}}_{\rm MLP} = 0.9501$. For the panel (f), we have $\Omega_{\rm op}^{\rm MP} = -0.34\epsilon$ and the same $\Omega_{\rm op}^{\rm MLP}$ (because the MLP optimal Rabi control is $\gamma$-independent), giving $\bar{\mathcal{F}}_{\rm MP} = 0.7986$ \red larger than \blk $\bar{\mathcal{F}}_{\rm MLP} = 0.7769$. Qubit trajectories in $x$ and $y$ coordinates are shown in (b),(d),(g), and (i), where solid curves are for optimal trajectories and dashed curves are for non-optimal ones (see text). Panels (c), (e), (h), and (j) show the corresponding histograms of the final states based on numerically simulating $10^4$ stochastic (unravelled) trajectories. The target state is shown as red crosses and red lines, with the grey bands showing the infidelity tolerance multiplied by 10 (for visibility).} 
\label{2}
\end{figure*}

We first show examples of the MP and MLP controls in Figure~\ref{2}, where, \blk from left to right, we plot average fidelities (a,f), qubit state trajectories and final-time (prepared) state histograms of the $x$ coordinate (b,c,g,h), and those of the $y$ coordinate (d,e,i,j). We choose to present results for two values of the dephasing rate (top row: weak $\gamma =0.1 \epsilon$, and bottom row: strong $\gamma = 0.8 \epsilon$) showing, respectively, when the MP and MLP approaches predict relatively similar optimal Rabi drive, and when they predict the optimal values differently. Since the MP method does not have a constraint on the final time, we choose to fix it at the optimal time $T_{\rm op} = 2.72 \epsilon^{-1}$ obtained from the MLP approach using Eq.~\eqref{TimeMLP}. We note that the range of parameters used in the numerical simulation here are motivated by actual parameters in superconducting qubit experiments~\cite{siddiqi2016}\red, where we assume a small leftover qubit's frequency $\epsilon$ (i.e., in the ${\rm MHz}$ range), the typical dephasing rate $\gamma \sim$ 0.01 - 0.2 MHz (i.e., the decoherence time in the range of 5 - 100 $\mu$s), and the Rabi drive in the range of $\Omega/2\pi \sim$ 0 - 5 MHz.\blk

The plots of average fidelities in Figure~\ref{2}(a,f) are simply to show that the MP Rabi controls always satisfy the maximum average fidelity values. In the weak dephasing regime, both MP and MLP trajectories are quite similar and the histograms of the final states are well concentrated around the target state. In the strong dephasing regime, when the MLP approach chooses a different Rabi drive from that of the MP, we can see their trajectories (coloured solid curves) diverge from each other. The coloured dashed curves are to show the non-optimal trajectories, where we swap the roles of optimal values, i.e., using the MP Rabi control in the MLP dynamical equation and vice versa.

An interesting feature arises in the strong dephasing regime (bottom row of Figure~\ref{2}), where we can see in the panel (h) that the histogram of the final states from the MLP approach (magenta histogram) does cover the target state (magenta line with grey band) much better than the histogram from the MP one (cyan histogram). That is, using the MP Rabi control, the state would not be able to reach the target state within an acceptable infidelity tolerance. In this work, we choose the infidelity tolerance based on the size of our trajectory data (see Appendix~\ref{app-fluc} for the full analysis), which gives $\delta = 0.005$ for the size of $10^4$ trajectories.
We note that, even though the MP histogram in the panel (j) for the $y$-coordinate covers the target value $y_{\rm F}$, those MP final states do not concentrate around the target state because of the deviation from the target value $x_{\rm F}$ in the $x$-coordinates.

\subsection{Comparison of success rates for various target states}

\begin{figure}
\centering
\includegraphics[width=14cm]{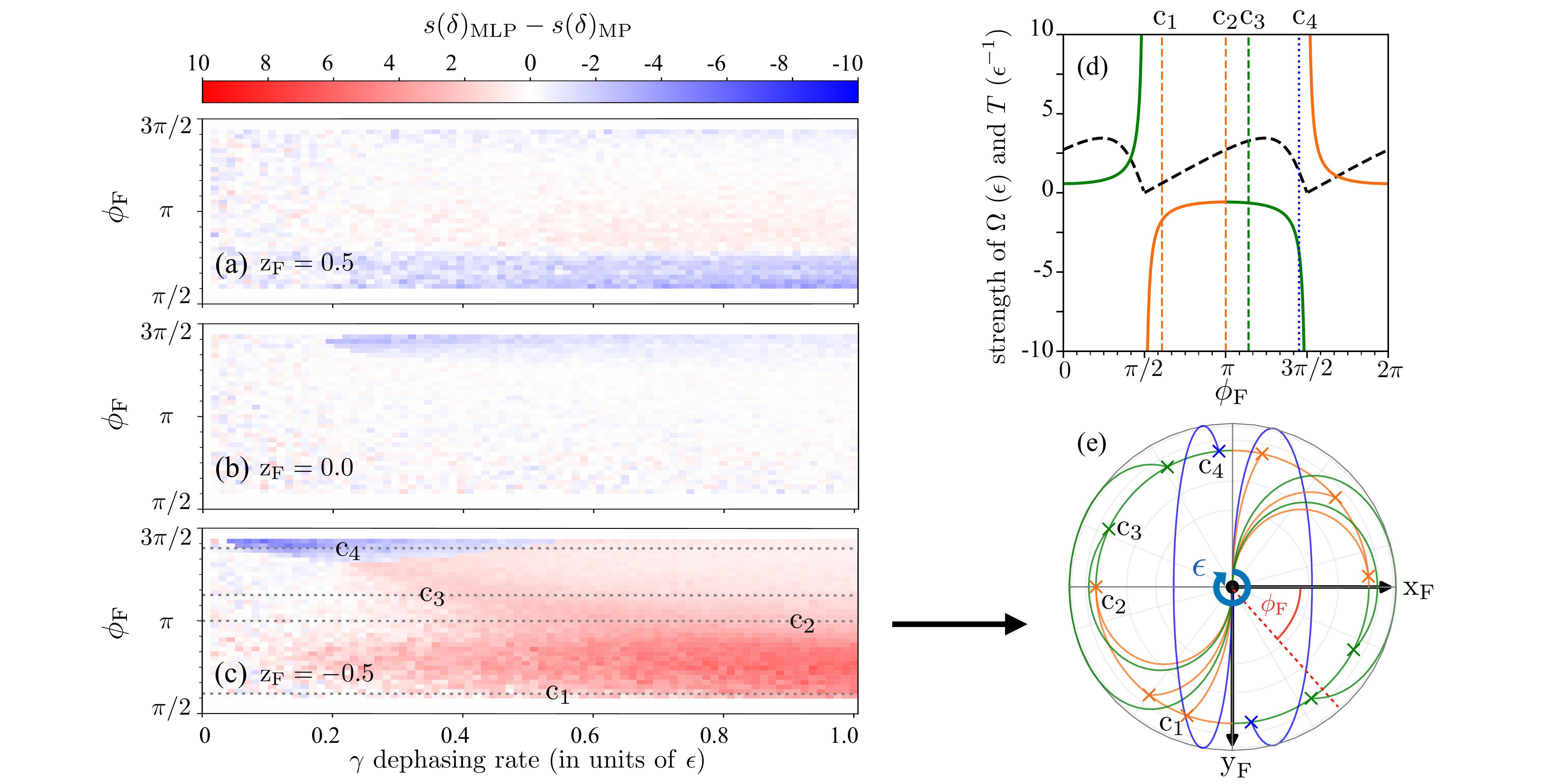}
\caption{Comparison of the success rates between the MLP and MP approaches, using the infidelity tolerance $\delta =0.005$, for various target states and values of the dephasing rate. The contour plots in (a), (b), and (c) show the difference: $s(\delta)_{\rm MLP} - s(\delta)_{\rm MP}$ for $z_{\rm F}=0.5$, $z_{\rm F}=0.0$, and $z_{\rm F}=-0.5$, respectively, where $\phi_{\rm F} \in [\pi/2,3\pi/2]$. To analyse the results in (c) further, we show in (d) the optimal Rabi control (solid coloured curves) and the MLP optimal time (dashed black curves) and show in (e) the MLP trajectories on the Bloch sphere projection, for the targets in the $z_{\rm F}=-0.5$ plane. In panel (e), given the initial ground state at the midpoint (black dot) of the sphere, the MLP paths can be categorized into: the short paths (orange) travelling directly from the bottom of the sphere to the targets (orange crosses) at the $z_{\rm F} = -0.5$ plane, the medium paths (green) travelling to the targets by slightly detouring over the $z_{\rm F} = -0.5$ plane, and the long paths (blue) travelling all the way around the $x$ axis and back to the target states. The colour code (orange, green, blue) is organized such that one can read off the Rabi drive and time from the plot in (d). (see text for the labels $c_1$, $c_2$, $c_3$, and $c_4$)}
\label{3}
\end{figure}

The histograms of the final states, shown in the previous subsection, can be used in computing the success rate Eq.~\eqref{eq-SR} of the control. Here, we explore the success rates for various target states ${\bm q}_{\rm F}$ as well as various values of the dephasing rates $\gamma$. In Figure~\ref{3}, we choose the target states in three planes: $z_{\rm F} = 0.5$, $z_{\rm F} = 0.0$, and $z_{\rm F} = -0.5$ and choose the Bloch sphere's azimuthal angle in the range: $\phi_{\rm F} \in [\pi/2, 3\pi/2]$ (there is a symmetry in $\pm x_{\rm F}$), where $x_{\rm F} = \sin(\arccos z_{\rm F})\cos\phi_{\rm F}$ and $y_{\rm F} = \sin(\arccos z_{\rm F})\sin\phi_{\rm F}$. We also choose the range of dephasing rates: $\gamma \in [0, 1]$ (in units of $\epsilon$). The contour plots in Figure~\ref{3}(a,b,c) show the difference of the success rates: 
\be
s(\delta)_{\rm MLP} - s(\delta)_{\rm MP},
\ee
(in units of percentage), where each data point is from simulating $10^4$ random noise trajectories. The red-coloured region indicates positive differences, meaning that the MLP control performs better than the MP, whereas the blue-coloured region indicates negative differences. The white region means that the MLP and MP controls' performance are relatively comparable.

We find that when the target state is far from the initial state, e.g., when $z_{\rm F}=0.5$ (panel (a)) and $z_{\rm F}=0$ in Figure~\ref{3}(b), the difference in the success rate is within $<2\%$, shown as mostly white and faint red-blue regions, indicating that both MP and MLP controls perform equally well. However, the significant difference between the two approaches occurs when the target state is close to the initial state $z_{\rm F} = -0.5$, as shown in panel (c), where the dark red region ($> 6\%$) covers a wide range of the target states and noise dephasing rates. 

Let us look closely at the $z_{\rm F} = -0.5$ plane, where there is a big dark red region (where the MLP definitely wins) and a small dark blue region (where the MP wins). 
We can understand those cases better by looking at the MLP optimal Rabi control (solid orange and green curves in Figure~\ref{3} panel (d)), the MLP optimal time (dashed black curves in panel (d)), and the MLP trajectories in the Bloch sphere projection in the $x$-$y$ plane (panel (e)). 
We choose three different target states, marked as $c_1$, $c_2$, and $c_3$ in Figure~\ref{3}(c), and show their associated MLP trajectories in Figure~\ref{3}(e). For $c_1$ and $c_2$, the MLP trajectories are categorized as short paths (orange curves in (e)) travelling from the bottom of the sphere directly to the targets in the $z_{\rm F}=-0.5$ plane, with short travelling times. For these states, because of their short paths, the MLP and MP controls perform equally well in the low-dephasing $\gamma < 0.3$ regime. However, with a stronger dephasing (high $\gamma$), the MLP controls perform significantly better than the MP controls  because the final states of the MP paths start losing their purity from the dephasing effect in the Lindblad equation. 

For the small dark blue region, where the MP wins, marked by the target state $c_3$, the qubit trajectory (blue curve in Figure~\ref{3}(e)) has to travel from the bottom (initial state) all the way to the north pole of the sphere and back to the target state on the $z_{\rm F} =-0.5$ plane. This has to occur with a strong Rabi drive in order to compete with the $\epsilon$ rotation. In this case, we find that the MP control can win over the MLP control, only in the small region of the low dephasing rate ($\gamma < 0.3$), because the MP state does not suffer much from the purity reduction, and so the method chooses a really large Rabi drive for the state to rotate more than one round to get closer to the target state (see a more detailed analysis in Appendix \ref{app-discussSR}). For the strong dephasing case ($\gamma > 0.3$), the MLP controls still win.

\subsection{Comparison with multi-pulse controls from the MP approach}

\begin{figure*}
\centering
\includegraphics[width=13cm]{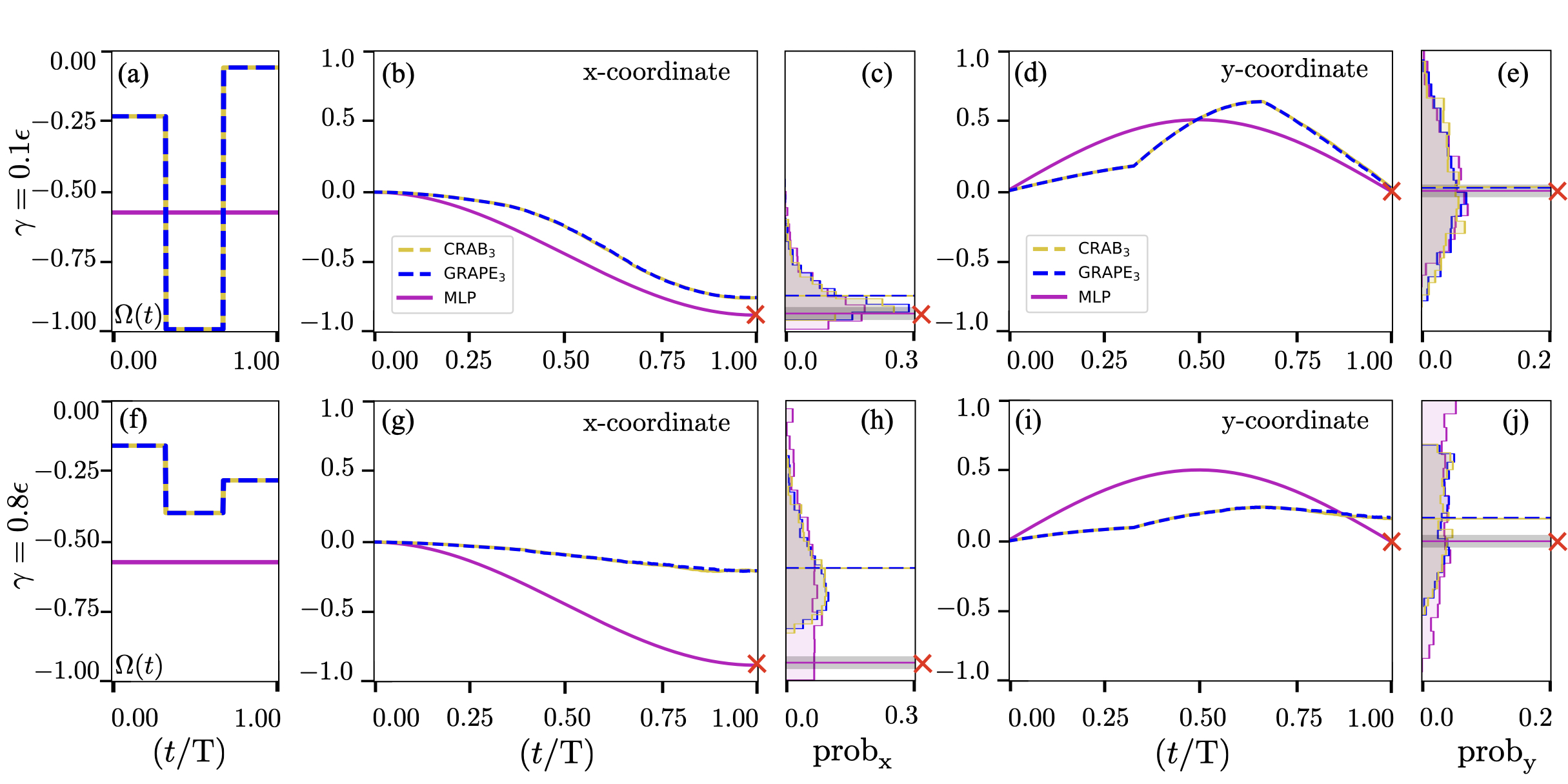}
\caption{Numerical results of the multi-pulse controls, comparing the MLP single pulse (magenta) derived analytically in Eqs.~\eqref{eq-solution}-\eqref{RabiTimeMLP} to the MP's $m=3$ multi pulses obtained using GRAPE$_m$ (blue dashed) and CRAB$_m$ (light yellow) algorithms. Panels (a) and (f) show the optimal Rabi drives from different approaches, while the rest of the panels shows corresponding qubit trajectories and final state histograms in both $x$ and $y$ coordinates, with the same parameters as those in Figure~\ref{2}. The MLP control gives $\Omega_{\rm op}^{\rm MLP} =  0.55$ in both the weak and strong dephasing cases. The MP's GRAPE$_3$ and CRAB$_3$ algorithms both give $\{\Omega_1, \Omega_2, \Omega_3\}=\{ -0.22,-1.00,-0.08\}$ for the low-strength dephasing and $\{-0.20,-0.30,-0.28\}$ for the high-strength dephasing. The interesting features can be seen in  panel (g) and (h), where the MLP control is the only approach that enables final states to lie in proximity to the target state, with a non-zero success rate (see texts and  Table~\ref{table1} for more detail).
}
\label{4}
\end{figure*} 

We have so far only shown the analyses of the MLP and MP control performances for the single-pulse, $\Omega_t=\Omega$, case. In this subsection, we allow the MP method to search for more complicated Rabi pulses, to see if they can improve the success rate over the MLP single-pulse case. We consider using the MP approach with a multi-pulse Rabi drive, which is a $m$ segment piecewise-constant function. To obtain an optimal MP control, we use GRAPE$_m$ and CRAB$_m$ algorithms to numerically search for multi-pulse Rabi drives that maximise the average fidelity between the final qubit state and the target state of Eq.~\eqref{eq-avefid-qubit}.

Numerical results are shown in Figure~\ref{4} for $m=3$ pulses. The reason we choose $m=3$ is such that: the number of pulse is not too small and the pulse is not too complex to be implemented in experiments. We also explored results for higher $m$'s and found that they did not offer much improvement to the average fidelity. For $m=3$, the Rabi drive is piece-wise constant in three equal time periods, i.e.,
 \begin{equation}
    \Omega_t = \left\{\begin{array}{lr}
        \Omega_1, &  t \in [0,\frac{1}{3}T)\\
        \Omega_2, & t \in [\frac{1}{3}T,\frac{2}{3}T)\\
        \Omega_3, &  t \in [\frac{2}{3}T,T)
        \end{array}\right. ,
 \end{equation}
where $T$ is the final time. We show in Figure~\ref{4}(a) and (f) the optimal Rabi drives from the MP multi-pulse and the MLP single-pulse approaches, for the weak and strong dephasing cases. Both MP's GRAPE$_3$ (dashed blue) and MP's CRAB$_3$ (light yellow) obtained the same pulses, indicating that the results are independent of the algorithms used. The MLP optimal Rabi single pulse (magenta) has the same value for both the weak and strong dephasing cases, as expected (it is independent of the dephasing rate). These controls result in different qubit trajectories shown in panels (b), (d), (g), and (i) of Figure~\ref{4}. Interesting features can be seen in the final state histograms. In the weak dephasing regime, the histograms are relatively similar for different control strategies. However, in the strong dephasing regime, the MLP single pulse is \textit{the only} control that results in the final state distribution containing the target state, as seen clearly in the panel (h) of Figure~\ref{4}.

We highlight our numerical results of the multi-pulse example in Table~\ref{table1}, where we show the calculated values of the average fidelities and the success rates, for different control approaches considered. As expected, the MP-based approaches (MP single pulse, MP's GRAPE, and MP's CRAB) give the highest average fidelities. However, considering the new quality metric (success rates), the MLP controls perform reasonably well in the low-dephasing regime (losing slightly to the MP multi-pulse controls) but significantly better than those MP's controls in the high-dephasing regime. The MP controls could not even make the qubit reach the target state with such a high value of the average fidelity.


\begin{table}[t]
	\caption{Summary of numerical results of the multi-pulse example shown in Figure~\ref{4}, comparing our proposed MLP controls with the traditional MP controls. We show the average fidelity \red ($\bar{\cal F}$ defined in Eq.~\eqref{eq-fidelave}) \blk and the success rates \red ($s$ defined in Eq.~\eqref{eq-SR})\blk computed from the final state histograms in Figure~\ref{4} using the infidelity tolerance $\delta = 0.01$ and $\delta = 0.005$.}\label{table1}
\centering
\begin{tabular}{ |p{2cm}||p{2cm}|p{2cm}|p{2cm}|  }

	\hline
	\multicolumn{4}{|c|}{Low-strength dephasing $\gamma=0.1\epsilon$} \\
	\hline
	Method&$\bar{\mathcal{F}}$&$s(\delta=0.01)$ &$s(\delta=0.005)$\\
	\hline
	MLP$_1$  & 0.947 & 22    &13\\
	MP$_1$ &0.950&   23 & 14   \\
	GRAPE$_3$ &  $\boldsymbol{0.954}$&$\boldsymbol{33} $& 20\\
	CRAB$_3$   &  $\boldsymbol{0.954}$ &32 & $\boldsymbol{21}$\\
	\hline
\end{tabular}
\end{table}
\begin{table}[t]
\centering
\begin{tabular}{ |p{2cm}||p{2cm}|p{2cm}|p{2cm}|  }
	\hline
	\multicolumn{4}{|c|}{High-strength dephasing $\gamma=0.8\epsilon$} \\
	\hline
	Method&$\bar{\mathcal{F}}$& $s(\delta=0.01)$ &$s(\delta=0.005)$\\
	\hline
	MLP$_1$&   0.773   & $\boldsymbol{4.6}$ &$\boldsymbol{2.2}$\\
	MP$_1$&0.792&   0 & 0   \\
	GRAPE$_3$&  $\boldsymbol{0.804}$& 0& 0 \\
	CRAB$_3$&  $\boldsymbol{0.804}$  & 0& 0\\
	\hline
\end{tabular}
\end{table}
\section{Conclusions}\label{sec6}
We have proposed the use of most-likely paths for controlling quantum systems in noisy environments, with a particular task to prepare the quantum systems at desired target states. Instead of the conventional mean-path (MP) approach using the Lindblad evolution as a representative of the controlled quantum system's state, in this work, we considered unravelling the Lindblad evolution to quantum trajectories for all possible environmental-noise realizations and searching for control functions associated with the most-likely noise path (MLP). In other words, we adopted the quantum evolution associated with the most-likely noise as a representative of the controlled system's state.

We applied both MP and MLP approaches to the example of qubit-state preparation, under the pure-dephasing noise environment, where only the qubit's Rabi drive can be controlled. Using the new state quality metric, the success rate, defined as a success probability that the controlled quantum state lies near the target within some fidelity error threshold, we found that the MLP controls resulted in higher success rates than the MP control in most cases, especially in the regime where the noise significantly influences the qubit (a strong dephasing rate). Even when the MP control was allowed to have multi-pulse Rabi drives searched numerically with GRAPE and CRAB algorithms, the MLP single-pulse could still perform better in the strong dephasing regime. This can be understood that, as the environmental dephasing rate increases, the Lindblad solution will suffer from the decoherence effect (e.g., losing its purity), making the standard Lindblad solution a poor representative of the controlled quantum system to reach a pure target state. We note that the single-qubit control example only serves as a proof of concept, where more sophisticated models will need to be investigated further in order to fully explore the use of most-likely paths in quantum control. 

\backmatter


\section*{Declarations}

\bmhead{Funding}

		This research has received funding support from the NSRF via the Program Management Unit for Human Resources and Institutional Development, Research and Innovation, grant number B39G670018 and by Thailand Science Research and Innovation Fund Chulalongkorn University IND66230005. We also acknowledge the National Science and Technology Development Agency, National e-Science Infrastructure Consortium, Chulalongkorn University and the Chulalongkorn Academic Advancement into Its 2nd Century Project (Thailand) for providing computing infrastructure that has contributed to the research results reported within this work.
		
\bmhead{Author contribution}
All authors contributed to the study conception and design. Material preparation, data collection and analysis were performed by Wirawat Kokaew. The first draft of the manuscript was written by Wirawat Kokaew, and Areeya Chantasri and Thiparat Chotibut commented on previous versions of the manuscript. All authors read and approved the final manuscript.

\begin{appendices}

\section{Deriving the Lindblad master equation}\label{app-open}


An open quantum system is defined as a quantum system $S$ that is coupled to another system $B$, known as a bath (or environment), which is typically assumed to be extremely large in comparison to the system $S$. Although the combined total system, $S+B$, is assumed to be a closed system, evolving unitarily with their Hamiltonian. Because of the system-environment interactions, the state of the system $S$ will no longer evolve unitarily as a closed system, but rather with some additional decoherence effects or measurement backactions. Let us begin with a combined system described by the total Hamiltonian similar to Eq.~\eqref{eq-hamtot} in the main text,
\be\label{eq:hamagain}
\hat{H}_{\rm{SB}} = \hat{H}_{\rm{S}}\otimes \hat{I}_{\rm B} + g \hat S \otimes \hat B(t),
\ee
where $\hat{H}_{\rm{S}}$ is a system-only Hamiltonian and $g \hat S \otimes \hat B(t)$ is the interaction Hamiltonian describing the interaction between the system and the bath. We note that the bath-only Hamiltonian is not of interest and can be removed by using the interaction frame with the bath's Hamiltonian and that the identity operator $\hat{I}_{\rm B}$ will be dropped later for convenience. Let us follow the notation in the main text and use $\varrho(t)$ as a density operator for the combined system. The dynamics of the combined system is described by the von-Neumann equation, 
\be\label{eq:von}
\dot{\varrho}(t)=-i[\hat{H}_{\rm{SB}},\varrho(t)],
\ee
This equation can be solved via a perturbative expansion if the interaction between the system and bath is assumed weak enough. We can integrate Eq.~(\ref{eq:von}) and then substitute the solution back into itself to yield
\be\label{eq:diffintegro}
\dot{{\varrho}}(t)=-i[{\hat H}_{\rm SB},{\varrho}(0)]-\int_0^tdt'[{\hat H}_{\rm SB}(t),[{\hat H}_{\rm SB}(t'),{\varrho}(t')]].
\ee
In order solve this equation, we make the following approximations. We assume that at $t=0$ there are no correlations between the system and the bath, so the initial state of the combined system can be factorized as ${\varrho}(0)={\rho}(0)\otimes{\rho}_{\rm B}(0)$. At any later time, the correlations may arise because of the interaction between the system and the bath. However, we also assume that the interaction is very weak and the bath is extremely large. Consequently, at any time $t$, the bath's state is not affected much, i.e., the state of the combined system at time $t$ is given by
\be
{\varrho}(t)\approx {\rho}(t)\otimes{\rho}_{\rm B}(0).
\ee
This is called the Born approximation or the weak-coupling approximation. Moreover, our interaction Hamiltonian Eq.~\eqref{eq:hamagain} has the first part that only acts on the system Hilbert space. Therefore, the system-only evolution given by tracing out the bath is,
\begin{align}\label{eq:master2nd}
\dot{\rho}(t) &= \Tr_{\rm B}\{ \dot\varrho(t) \}  \\
&=-i[\hat{H}_{\rm{S}},\rho(t)] - g^2 \int_0^t dt' \Tr_{\rm B}\{ [\hat S \otimes \hat B(t),[\hat S \otimes \hat B(t'),{\rho}(t')\otimes{\rho}_{\rm B}(0)]] \} \nonumber,
\end{align}
where we have assumed $\Tr_{\rm B}\{ [\hat H_{\rm SB}, \varrho(0)]\} = 0$ as the initial condition for simplicity. The time integral in Eq.~\eqref{eq:master2nd} is still not easy to solve as the present evolution ${\rho}(t)$ depends on the past ${\rho}(t')$. In order to further simplify the integration, we make the third significant assumption called the Markov assumption, where in essence  ${\rho}(t')$ is replaced by ${\rho}(t)$ and the limit of time integration can be extended to negative infinity.
Expanding terms in Eq.~\eqref{eq:master2nd}, we find
\begin{align}
\label{eq:fullform}
\dot{{\rho}}&(t)=-i[\hat{H}_{\rm{S}},\rho(t)] + 2 g^2 \int_{-\infty}^t dt' \left\{ \hat S \rho(t) \hat S - \frac{1}{2} (\hat S \hat S \rho(t) + \rho(t) \hat S \hat S) \right\} C(t,t') .
\end{align}
We have defined the correlation functions as 
\begin{align}
C_{ij}(t,t')\, & \equiv\Tr_{\rm B}\{{\hat B}(t){\hat B}(t'){\rho}_{\rm B}(0)\} \nonumber\\
&= \Tr_{\rm B}\{{\hat B}(t'){\hat B}(t)\rho_{\rm B}(0)\} = \kappa \delta(t-t'),
\end{align}
where we have assumed the delta-correlated noise, justifying the Markov assumption as the bath's correlation times are much shorter compared to the system's time, where $\kappa$ is a constant spectral density.
Therefore, we obtain the Lindblad equation,
\begin{eqnarray}
\dot{{\rho}}(t)=-i[\hat{H}_{\rm S}, {\rho}(t)]+\gamma\mathcal{D}[\hat{S}]{\rho}(t),
\end{eqnarray}
where $\gamma\equiv2g^2\kappa$ and $\mathcal{D}[\hat{S}]\bullet\equiv\hat{S}\bullet \hat{S}^\dagger -(\hat{S}^\dagger \hat{S}\bullet+\bullet \hat{S}^\dagger \hat{S})/2$, as used in Eq.~\eqref{eq-decoh} in the main text.

\section{Deriving the optimal MLP Rabi control and time}
In the main text, we have shown the results of the optimal Rabi control and its associated optimal time for the MLP approach. Here, we show the full derivation in two different ways. One is by solving the six differential equations Eqs.~\eqref{eq-6ODEs} with two constraints Eqs.~\eqref{eq-constraint}. The other one is more intuitive and is solved via a geometrical interpretation of a rotation on the qubit's Bloch sphere.

\subsection{Derivation via differential equations}\label{app-diff}
Let us begin with the time-continuous version of the equations describing the dynamics of the most-likely path, which are the six ODEs  
\begin{subequations}\label{eq:ODE}
\begin{eqnarray}
\dot{x}=- y\tilde{\epsilon},\qquad
&\dot{p}_{x}=- p_{y}\tilde{\epsilon},\\
\dot{y}=x\tilde{\epsilon} +z\Omega
,&\qquad
\dot{p}_{y}=p_{x}\tilde{\epsilon} +p_{z}\Omega
,\\
\dot{z}= - y\Omega,\qquad
&\dot{p}_{z}= p_{y}\Omega.
\end{eqnarray}
\end{subequations}
and the two constraints
\begin{subequations}
\begin{eqnarray}
\xi&=&-2g\kappa p_{x}y +2g\kappa p_{y}x\label{eq:c1},\\
p_{y}z&=&p_{z}y\label{eq:c2},
\end{eqnarray}
\end{subequations}
where we have dropped the time argument from all variables for simplicity. First, we consider taking the time derivative of Eq.~(\ref{eq:c2}),
\be\label{eq:c4}
	\dot p_yz+\dot zp_y=\dot p_zy+\dot yp_z.
\ee
We can substitute the relations of $\dot{p}_y, \dot{z}, \dot{p}_z$, and $\dot{y}$ from Eq.~(\ref{eq:ODE}) into Eq.~(\ref{eq:c4}) and obtain
\begin{eqnarray}\label{eq:c3}
	\tilde{\epsilon} p_x z+\Omega p_z z-\Omega p_y y &=-\Omega p_yy+\tilde{\epsilon} p_z x+\Omega p_z z,
\end{eqnarray}
where most terms cancel out and we get $p_xz=p_zx$ as shown in the main text. We can use this new relation in Eqs.~(\ref{eq:c1}) and (\ref{eq:c2}), where we obtain the zero noise solution, $\xi=0$. Then, we can reduce the six equations of motion in Eqs.~(\ref{eq:ODE}) to just three:
\begin{subequations}
\begin{eqnarray}
		\dot x&=& -\epsilon y,\\
		\dot y&=& \epsilon x+\Omega z,\\
		\dot z&=&-\Omega y.
\end{eqnarray}
\end{subequations}
Using the boundary condition: $\{x(0),y(0),z(0)\}$$=$$\{x_{\rm I},y_{\rm I},z_{\rm I}\}$ and $\{x(T),y(T),z(T)\}=\{x_{\rm F},y_{\rm F},z_{\rm F}\}$, we find that the general solution becomes
\begin{subequations}\label{eq:sol}
\begin{eqnarray}
		x_{\rm F}&= &\alpha_1\frac{\epsilon}{\Omega}\cos(\omega T)+\alpha_2\frac{\epsilon}{\Omega}\sin(\omega T)-\alpha_3\frac{\Omega}{\epsilon}\\
		y_{\rm F}&=& \alpha_1\frac{\omega}{\Omega}\sin(\omega T)-\alpha_2\frac{\omega}{\Omega}\cos(\omega T),\\
		z_{\rm F}&=&\alpha_1\cos(\omega T)+\alpha_2\sin(\omega T)+\alpha_3.
\end{eqnarray}
\end{subequations}
where we have used $\omega\equiv\sqrt{\Omega^2+\epsilon^2}$ and defined
\begin{subequations}
\begin{eqnarray}
 \alpha_1&=&\frac{1}{\omega^2}(\Omega^2z_{\rm{I}}+\epsilon\Omega x_{\rm{I}}),\\ \alpha_2&=&-\frac{1}{\omega}y_{\rm{I}}(\Omega),\\
 \alpha_3&=&\frac{1}{\omega^2}(\epsilon^2z_{\rm{I}}-\epsilon\Omega x_{\rm{I}}).
\end{eqnarray}
\end{subequations}
Solving Eqs.~(\ref{eq:sol}) for the Rabi oscillation and time, we get
\begin{align}\label{eq:RabiMLP}
\Omega^{\rm MLP}_{\rm op} &= \epsilon\left(\frac{z_{\rm{F}}-z_{\rm{I}}}{x_{\rm{F}}-x_{\rm{I}}}\right),\\
		\label{eq:TimeMLP}
T^{\rm MLP}_{\rm op}\!\!=&\frac{1}{\omega}\cos^{-1}\left(\frac{\Omega\omega x_{\rm{F}} \alpha_1-\Omega\epsilon y_{\rm{F}}\alpha_2+\alpha_1\alpha_2\Omega^2\omega/\epsilon}{\epsilon\omega(\alpha_1^2+\alpha_2^2)}\right),
\end{align}
which are for any initial and final states. However, for simplicity, we assume that the initial state for the quantum state preparation control is the qubit's ground state, i.e., we set $\{x_{\rm{I}},y_{\rm{I}},z_{\rm{I}} \}=\{ 0,0,-1\}$ and obtain
\begin{equation}
	\Omega^{\rm{MLP}}_{\rm{op}} = \epsilon\frac{z_{\rm{F}}+1}{x_{\rm{F}}},
\end{equation}
and 
\begin{equation}
	T^{\rm{MLP}}_{\rm{op}}=\frac{1}{\omega}\arccos \Big(1-\frac{x_{\rm{F}}\omega^2}{\epsilon\,  \Omega^{\rm{MLP}}_{\rm{op}}} \Big),
\end{equation}
which agree with Eqs.~\eqref{RabiTimeMLP} in the main text for $\{ x_{\rm{I}},y_{\rm{I}},z_{\rm{I}}\}=\{0,0,-1\}$.
We also note that, since we have written the solution in terms of the arccos function, the optimal time solution above will only be valid for  target states in the second and forth quadrants of the $x$-$y$ plane, denoted by $T_{2,4}$. For target states in the first and third quadrants, the optimal time solution can be obtained from $T_{1,3}=2\pi-T_{2,4}$.

\subsection{Derivation via Bloch-sphere geometrical approaches}\label{app-derive}
\begin{figure}
	\centering
	\includegraphics[width=12cm]{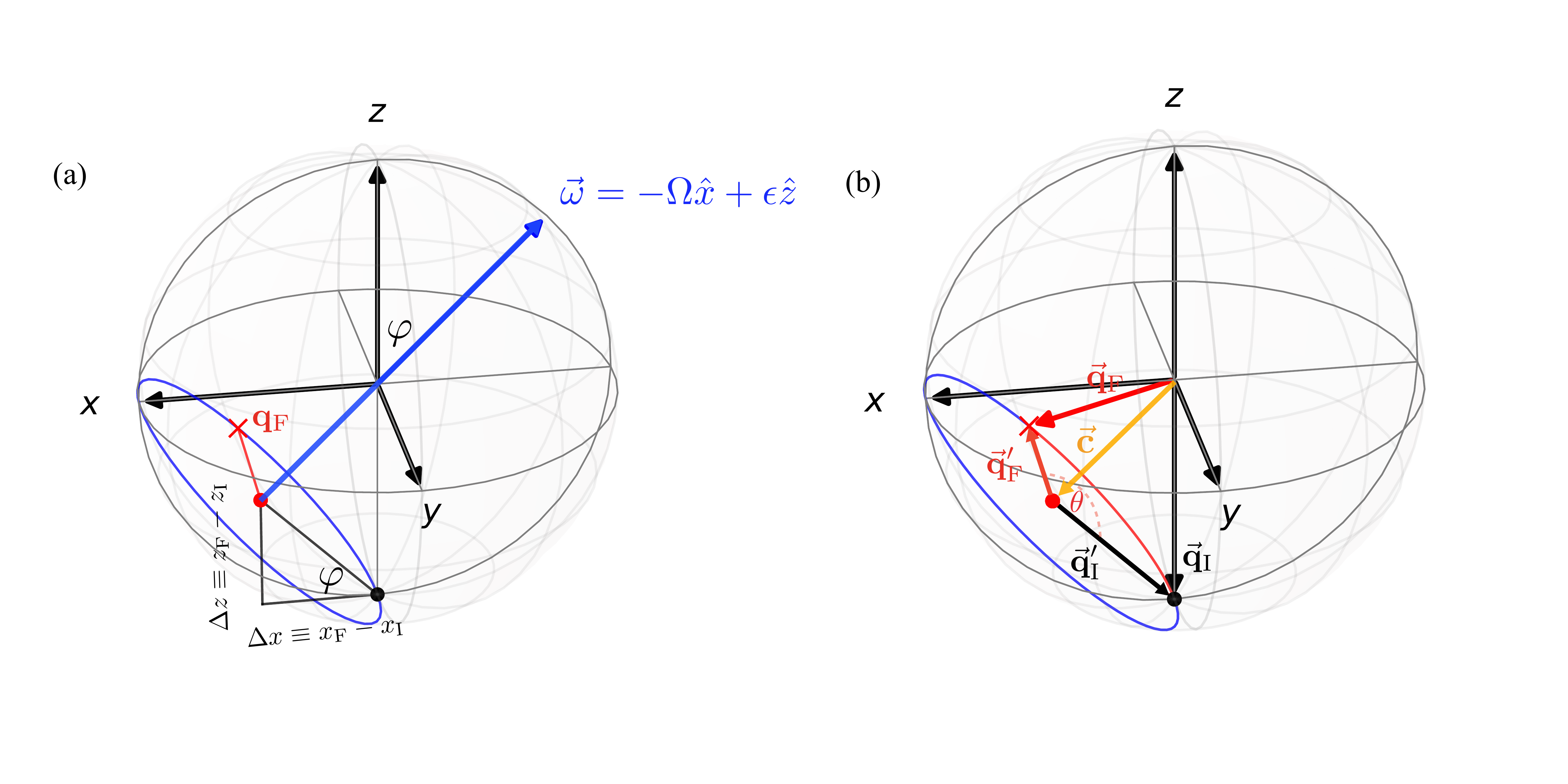}
	\caption{Schematic diagrams for deriving the MLP optimal solution using a geometrical approach. (a) The rotation axis, $\vec{\omega}$, is shown as a blue vector with an angle $\varphi$ from the $z$ axis, where the circle on the plane of optimal rotation (blue curve) shows the only rotation path from the ground state (black dot) to the target state (red cross). Interestingly, the axis of rotation $\vec{\omega}$ can be found from projecting the target state onto the $x$-$z$ plane (shown as the red dot) and that point will become the center of the circle on the plane of optimal rotation (blue). (b) The rotation angle, $\theta$, can be used to compute the optimal time via the relation $T=\theta/\omega$. To find this angle, we solve for the two vectors, $\vec{\boldsymbol{q}}'_{\rm I}$ and $\vec{\boldsymbol{q}}'_{\rm F}$ and compute their dot product (see the explanation in the paragraph containing Eq.~\eqref{eq:theta}).}
	\label{fig:c1}
\end{figure}

As the MLP approach suggested, the optimal control can be obtained from the zero-noise solution $\xi(t)=0$. Substituting the zero-noise realization into Eq.~\eqref{H} in the main text, the qubit's dynamics becomes a simple unitary rotation,
\be
	\hat{H}(t)= \frac{{\epsilon}}{2} \hat \sigma_z -\frac{\Omega}{2}\hat\sigma_x,
\ee
where we have already assumed the single pulse Rabi drive, i.e., $\Omega(t) = \Omega$. Since the $\epsilon$ is fixed, the task to find an optimal control is now reduced to finding $\Omega(t)$ that can bring the initial state ${\bm q}_I = \{ 0,0,-1\}$ to a desired target state ${\bm q}_{\rm F} = \{ x_{\rm F}, y_{\rm F}, z_{\rm F}\}$. 

Let us explore the use of a geometrical approach illustrated in Figure~\ref{fig:c1} to solve for the optimal Rabi drive and time. Given the unitary rotation above, the rotation axis is given by $\vec{\omega}=\epsilon\hat{z} - \Omega \hat{x}$, which is a combination of a rotation around the $\hat{z}$ axis with an angular speed $\epsilon$ and around the -$\hat{x}$ axis with a controllable angular speed $\Omega$. We can then draw the plane of rotation, which embeds the blue ``circle on the plane of optimal rotation" living on the Bloch sphere in Figure~\ref{fig:c1}(a) and (b), that represents the most-likely path to rotate the initial state to the target state. Therefore, the goal is to solve for the appropriate axis of rotation $\vec{\omega}$, which will tell us the optimal $\Omega$, and then solve for the angular distance between the two boundary states in order to find the optimal time used to travel between the two.

Let us start with finding the axis of rotation. From Figure~\ref{fig:c1}(a), we can see that $\varphi$ is the angle between the $\hat{z}$ axis and $\vec{\omega}$, as well as the angle of elevation from the initial to the target states in the $x$-$z$ plane (noting that the center (red dot) of the circle of the sphere is on the $x$-$z$ plane). The latter gives the tangent as the distance ratio: $\tan\varphi=(z_{\rm F} - z_{\rm I})/(x_{\rm F} - x_{\rm I})$. We can therefore find the optimal $\Omega$ by looking at the ratio of the dot products
\be
\frac{\hat{x}\cdot\vec{\omega}}{\hat{z}\cdot\vec{\omega}}=\frac{\omega\cos(\pi/2 + \varphi)}{\omega\cos(\varphi)},
\ee
where $\omega = \norm{\vec{\omega}} = \sqrt{\Omega^2 + \epsilon^2}$, which gives the optimal Rabi drive,
\begin{eqnarray}\label{eq-ca}
\Omega=\epsilon\tan\varphi = \epsilon \left( \frac{z_{\rm F}- z_{\rm I}}{x_{\rm F}- x_{\rm I}} \right) \label{eq-cb},
\end{eqnarray}
as in Eq.~\eqref{eq:RabiMLP} and in the main text Eq.~\eqref{RabiMLP}.

In order to find the rotation angle or the angular distance between the initial and the target states, denoted by $\theta$, we need to consider the angle between the two vectors, $\vec{\boldsymbol{q}}'_{\rm I}$ and $\vec{\boldsymbol{q}}'_{\rm F}$, shown in 
Fig.~\ref{fig:c1}(b). The two vectors are from the center point (red dot) of the circle on the plane of optimal rotation to the initial state (black dot) and the final state (red cross), respectively. We find the relations between these vectors and the vectors to the initial and final states from $\vec{\boldsymbol{q}}'_{\rm I}=\vec{\boldsymbol{q}}_{\rm I}-\vec{\boldsymbol{c}}$ and $\vec{\boldsymbol{q}}'_{\rm F}=\vec{\boldsymbol{q}}_{\rm F}-\vec{\boldsymbol{c}}$. We then calculate the $\theta$ using the dot product,
\begin{eqnarray}\label{eq:theta}
\theta&=&\arccos\Big(\frac{\vec{\boldsymbol{q}}'_{\rm I}\cdot\vec{\boldsymbol{q}}'_{\rm F}}{\norm{\vec{\boldsymbol{q}}'_{\rm I}}\norm{\vec{\boldsymbol{q}}'_{\rm F}}}\Big)\\
&=&\arccos\Big(\frac{-z_{\rm F}\sin\varphi-x_{\rm F}\cos\varphi}{\{1+\cos^2\varphi+2\cos\varphi(z_{\rm F}\cos\varphi-x_{\rm F}\sin\varphi)\}^{1/2}} \Big)\nonumber,
\end{eqnarray}
where we can substitute $\sin(\varphi)=\Omega/\omega$ and $\cos(\varphi)=\epsilon/\omega$ and obtain the optimal time using $T = \theta /\omega$ as shown in Eq.~\eqref{eq:TimeMLP} and in the main text Eq.~\eqref{TimeMLP}.

\section{Determining an appropriate infidelity tolerance}\label{app-fluc}

\begin{figure*}
	\centering
	\includegraphics[width=13cm]{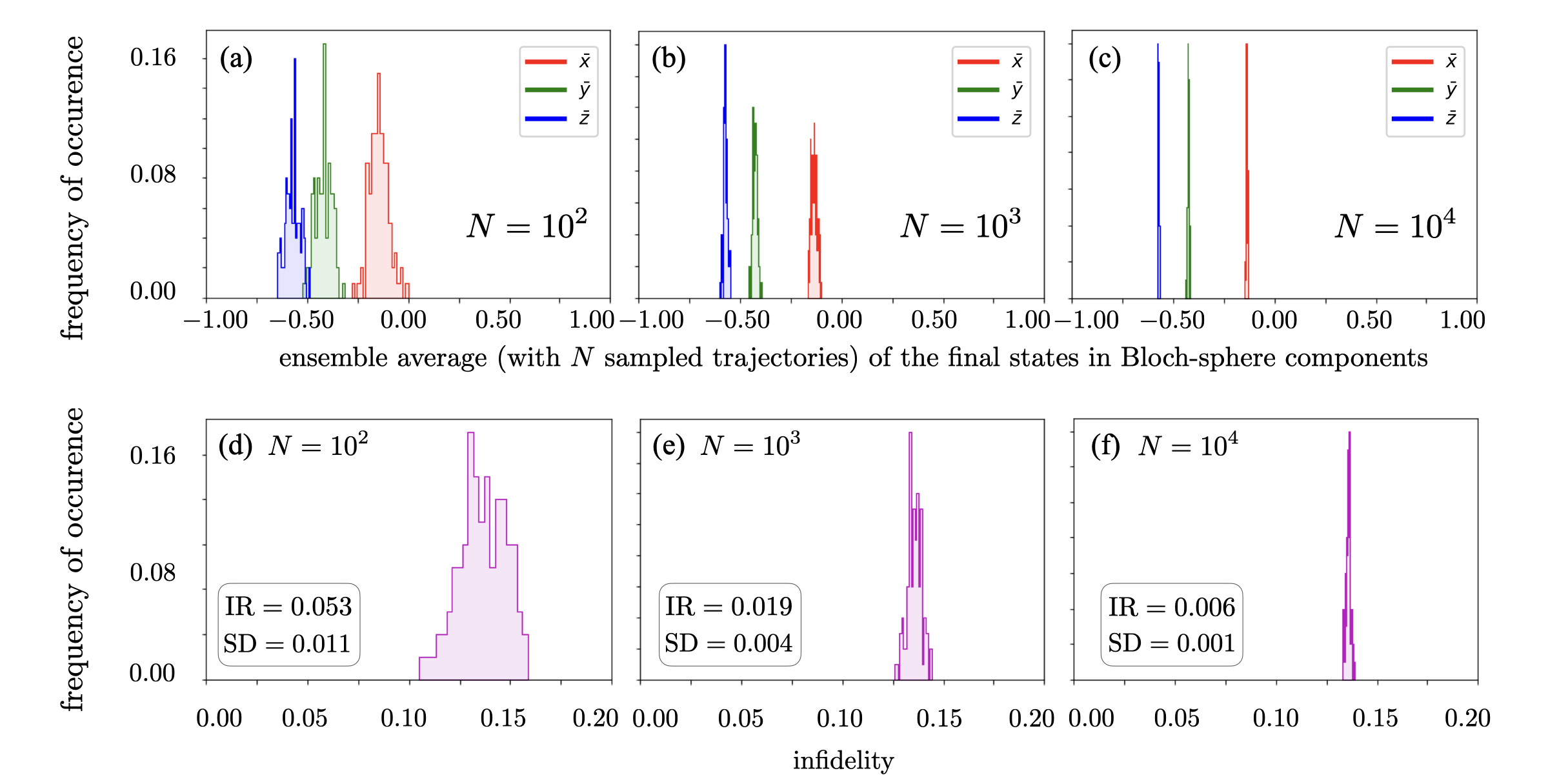}
	\caption{Top row: distributions of mean values (total of 100 values) of qubit's final states in the Bloch sphere components $x$ (red), $y$ (green) and $z$ (blue), for averaged trajectory-ensemble sizes of $N = 10^2$, $N = 10^3$, and $N = 10^4$ in the panel (a), (b), and (c) respectively. Bottom row:  distributions of average infidelities to a target state, ${\bm q}_{\rm F} = \{-0.16,-0.58,-0.8\}$, computed from the $x,y,z$ coordinates in the top row. The infidelity range (IR) is the absolute difference between the maximum and minimum values in the distributions and the standard deviation (SD) is computed from the distributions.}
	\label{fig:d}
\end{figure*}

\begin{figure*}
	\centering
	\includegraphics[width=13cm]{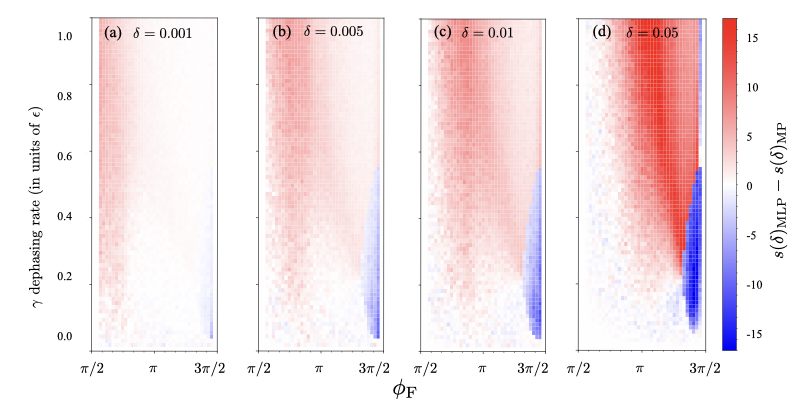}
	\caption{The difference of success rates from MLP and MP approaches, using the exact same trajectory data as in Figure~\ref{3} in the main text, but with varying infidelity tolerances: $\delta = 0.001, 0.005, 0.01$, and $0.05$, in the panels (a), (b), (c), and (d), respectively.}
	\label{fig:e2}
\end{figure*}

This is to ask a question: How close should the final state and the target state be, such that we can call it a success? In an ideal case, the success would be that the final state is exactly at the target. However, with a finite number of stochastic trajectories (a finite-size ensemble), the occurrence of the exact agreement is highly unlikely. Therefore, we need to introduce the infidelity tolerance in order to reasonably evaluate the success rate of a finite-size ensemble (as in Eq.~\eqref{eq-SR}). 

In order to arrive at a criterion of the infidelity tolerance, we consider finite-size fluctuations as a proxy for acceptable errors. For an ensemble of size $N$, if finite size fluctuations of the mean (set at the target state) follows the central limit theorem (CLT), one could reasonably argue that a state within a $1/\sqrt{N}$ relative error from the target is a success, as the error would stem from a finite-size sampling. In Figure~\ref{fig:d}, we show how the distributions of 100 mean (average) values  evolve as the number of trajectories (to be averaged) grows from $N = 10^2$, to $10^3$ and $10^4$. Figure~\ref{fig:d}(top) suggests that the distributions of the mean possess a CLT-like characteristic as $N$ increases; in fact, the standard deviation of these distributions roughly scales as $1/\sqrt{N}$. Figure~\ref{fig:d}(bottom) shows that the distributions of the average infidelity to a target state, $1- \bar{\cal F}(\rho_T, \rho_{F})$, have a finite support and that the fluctuations also shrinks as $N$ grows. Since we chose $N = N_{\rm tot} = 10^4$ in the main text, we set the value of infidelity tolerance from the statistics presented in Figure~\ref{fig:d}(f), revealing a fluctuation range (i.e., the difference between the maximum and minimum values in the distribution) of $0.006$ and the standard deviation of $0.001$. This leads to choosing the simple numbers such as $\delta = 0.001$, $\delta = 0.005$ or $\delta = 0.01$ as our infidelity tolerance.

In Figure~\ref{fig:e2}, we show how the success rate results in Figure~\ref{3} in the main text can be affected by different infidelity tolerance, $\delta$: $0.001$, $0.005$, $0.01$, and $0.05$. Overall, the four plots have the same feature showing regions with red, blue, and white regions. When $\delta=0.001$ in (a), we can see that most regions are white, which means that we cannot distinguish the performance of the MP and MLP approaches, since finite-size fluctuations exceeding the infidelity tolerance renders both methods unsuccessful. On the other hand, if the infidelity tolerance is too big, $\delta=0.05$ as shown in (d), the resulting success rates can be misleading, as states further away from the target are mistakenly considered successful state preparations.\\\\


\section{Characteristics of control protocols in different regimes of dephasing rates}\label{app-discussSR}

\begin{figure}
	\centering
	\includegraphics[width=14cm]{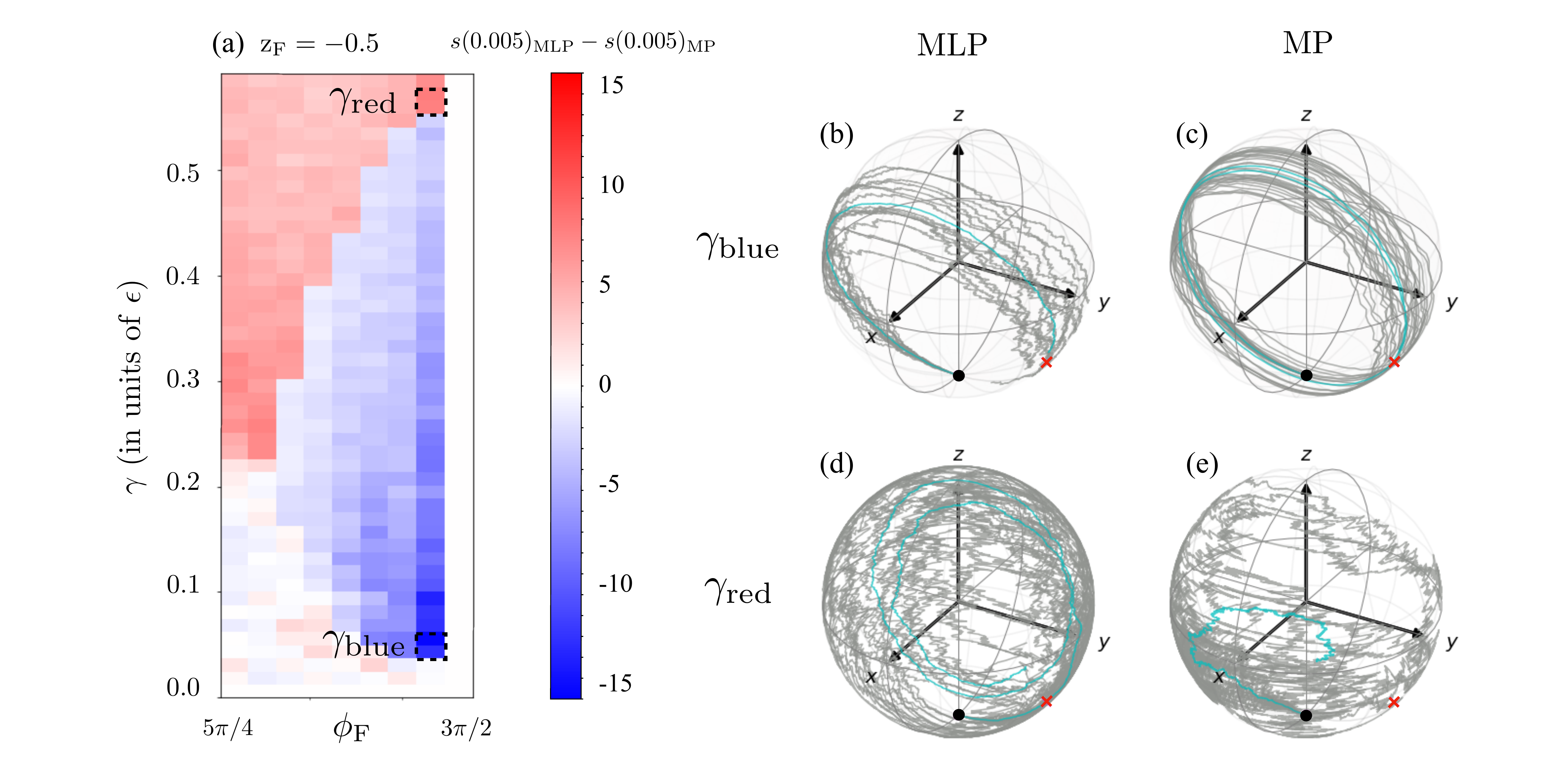}
	\caption{Further investigation of the characteristics of each control protocol in the blue and red regions of Figure~\ref{3}(c) in the main text. (a) the zoomed in contour plot of Figure~\ref{3}(c) in the main text, with two little dashed boxes showing two values of the dephasing rate denoted by $\gamma_{\rm blue}$ and $\gamma_{\rm red}$ that we investigate the trajectory characteristics. Each of the four Bloch sphere plots show 10 samples of the stochastic qubit trajectories (grey lines) and their average trajectories (cyan lines). (b) and (c) are the results of $\gamma_{\rm blue}$ using the MLP and MP Rabi controls, respectively. (d) and (e) are the results of $\gamma_{\rm red}$ using the MLP and MP Rabi controls, respectively.}
	\label{fig:e1}
\end{figure}

As we have shown in Figure~\ref{3}(c) of the main text, there are blue regions when the MP control yields a higher success rate than that of the MLP control. To understand what happens for those cases, we investigate the interesting region, $z_{\rm F}=-0.5$ and $\phi_{\rm F}=[\pi/4,\pi/2]$ (with $\delta=0.005$), as shown in Fig.~\ref{fig:e1}(a). We pick out two values of the dephasing rate indicated by the two little dashed boxes, denoted by $\gamma_{\rm blue}$ and $\gamma_{\rm red}$, for the ones sitting in the blue and the red regions, respectively.  We show the qubit trajectories using with the MLP and MP Rabi controls associated with $\gamma_{\rm blue}$ in Figs.~\ref{fig:e1}(b) and (c), respectively. The MLP approach chooses a slow Rabi drive which drives the qubit state from its initial state (black dot) to the target state (red cross), \red following the path similar to $c_4$ (blue curve) in Figure~\ref{3}\blk. However, the MP approach\red, which does not have to satisfy the optimal drive based on the geometric equation in Eq.~\eqref{eq-cb} and can choose any arbitrary size of the Rabi drive, instead prefers a large Rabi drive \red that further minimises the purity reduction (and thus maximises the fidelity to the target state)\blk, causing the state to rotate around the Bloch sphere more than one round in order to bring the average path (cyan line) close to the target state as much as possible.

In Figs.~\ref{fig:e1}(d) and (e), we also show qubit trajectories associated with the dephasing rate in the red region $\phi_{\rm red}$. We can clearly see that the trajectories are more strongly fluctuating than those in the low dephasing case of (b) and (c). In this case, it is not clear why the MLP approach exhibits a higher success rate than the MP one, but we can see that the average path (cyan line) from the MP approach in panel (e) suffers from the dephasing effect leading to a final state lying deep inside the Bloch sphere (small state purity) far from the actual target state.

\end{appendices}

\bibliography{ref}

\end{document}